\documentclass[useAMS,usenatbib,usegraphicx]{mn2e}
\usepackage{times}
\usepackage{natbib}
\title[Evidence of nearby supernovae affecting life on Earth]
{Evidence of nearby supernovae affecting life on Earth}
\author[H. Svensmark]{Henrik Svensmark$^{1}$\thanks{E-mail:
hsv@space.dtu.dk (Paper accepted by MNRAS 2012 March 17)} \\ $^1$ National Space Institute, Technical University of Denmark,
\\Juliane Marie Vej 30, 2100 Copenhagen {\O}, Denmark}
\begin{document}
\maketitle

\begin{abstract}
Observations of open star clusters in the solar neighborhood are used to calculate local supernova (SN) rates for the past 510 million years (Myr). Peaks in the SN rates match passages of the Sun through periods of locally increased cluster formation which could be caused by spiral arms of the Galaxy. A statistical analysis indicates that the Solar System has experienced many large short-term increases in the flux of Galactic cosmic rays (GCR) from nearby supernovae. The hypothesis that a high GCR flux should coincide with cold conditions on the Earth is borne out by comparing the general geological record of climate over the past 510 million years with the fluctuating local SN rates. Surprisingly a simple combination of tectonics (long-term changes in sea level) and astrophysical activity (SN rates) largely accounts for the observed variations in marine biodiversity over the past 510 Myr. An inverse correspondence between SN rates and carbon dioxide (CO$_2$) levels is discussed in terms of a possible drawdown of CO$_2$ by enhanced bioproductivity in oceans that are better fertilized in cold conditions - a hypothesis that is not contradicted by data on the relative abundance of the heavy isotope of carbon, $^{13}$C.
\end{abstract}


\section{INTRODUCTION}
\label{intro}
That life on Earth has always been subjected to strong influences from the cosmos has been among the main revelations in geology in recent decades. Headlines include the verification of the planetary Milankovitch effect as a pacesetter of glacial cycles, the realisation that life was unsustainable during a heavy bombardment of the young Earth (Hadean Eon), and the evidence that the Mesozoic Era of giant reptiles ended suddenly when an asteroid hit Mexico. Learning from such terrestrial examples, astrobiologists have wondered whether cosmic hazards may make some planetary systems unsuitable for life. For example, by analogy to the Goldilocks Zone of optimal stellar irradiation, \citet{Lineweaver2004} discuss a Galactic habitable zone in the Milky Way where one requirement is "an environment free of life-extinguishing supernovae".

Even though life in general has survived robustly on our planet for billions of years, the fossil evidence tells of continual changes among the inhabiting species in an ever-variable climate. Ninety years ago the astrophysicist \citet{Shapley1921} suggested that ice ages on the Earth might be due to the Solar System's encounters with gas clouds in the Milky Way. That idea was revived half a century later by \citet{McCrea1975Natur}, pursued by \citet{Talbot1977ApJ} and developed recently using better observations by \citet{Frisch2000AmSci}. But among some of those investigating risks associated with Earth's interaction with the interstellar medium, interest shifted to possible climatic and biological effects of radiation from supernovae exploding nearby. An "ultraviolet deluge" at the Earth's surface, due to formation of nitrogen oxides and consequent damage to the ozone layer, was proposed in 1974 by \citet{Ruderman1974Sci}. Others examined the idea \citep{Whitten1976Natur,Gehrels2003ApJ} and even suggested that supernovae could cause mass extinctions of living species \citep{Tucker1968Sci,Russell1971Natur,Reid1978Natur}.

Consideration of the effects of nearby supernovae (SNs) has recently focused on the influence of Galactic cosmic rays (GCR) generated by SN remnants. As discussed more historically in Sect. \ref{Sec6}, empirical evidence suggests that ionization of the air by GCR has influenced the terrestrial climate on time scales ranging from days \citep{Svensmark2009GeoRL} to billion of years \citep{Shaviv2003NewA}. Sufficiently energetic GCR primaries from SNs ($>$10 GeV) provoke showers of secondary particles in the atmosphere that include muons which dominate at the lowest altitudes. The hypothesis \citep{Svensmark2000PhRvL} is that ionization by the secondary particles helps to seed the formation of low clouds, by assisting the formation of aerosols (r $\approx$ 2-3 nm), some of which subsequently grow into cloud condensation nuclei (r larger than $\approx$ 50 nm). A high flux of GCR results in an increase in the number of cloud condensation nuclei which in turn increases the albedo of the clouds. As low clouds exert a cooling effect by increasing the Earth's albedo, high GCR fluxes imply low global temperatures, and vice versa. The chemical mechanism that promotes the creation of cloud condensation nuclei from sulphur compounds in the air has been verified in the laboratory \citep{Svensmark2007RSPSA,Enghoff2011GRL,Svensmark2012}, and observationally the whole chain from GCR, to aerosols, to clouds has been observed in connection with sudden solar coronal mass ejections on time scales of days \citep{Svensmark2009GeoRL,JSvensmark2012}.

The energetic GCR that ionize the lower atmosphere are only weakly influenced by variations in the geomagnetic field or by solar magnetic activity. Both cause low-altitude ionization rates to vary by ($\approx$10\%) in the course of a magnetic reversal or during a solar cycle. Over decades to millennia the GCR influx to the Solar System scarcely changes. On longer time scales, changes in GCR very much larger than those due to geomagnetic or solar activity occur as a result of variations in the rate of nearby SNs. Since the the main ionization in the Earth's lower atmosphere is caused by 10-20 GeV GCR, such energies will be implicitly assumed in the following.

\citet{Fields1999NewA} speculated that increased cloud cover due to GCR from a very close SN could cause a "cosmic ray winter". A more comprehensive scenario from Shaviv \citep{Shaviv2002PRL,Shaviv2003NewA,Shaviv2004GSA} linked icy episodes on the Earth during the 542 Myr of the Phanerozoic Eon to the Solar System's encounters with spiral arms of the Milky Way as it orbited around the Galactic centre. Shaviv attributed the climatic effect to enhanced GCR, as did \citet{Marcos2004NewA} in a study that used local star formation rates as a proxy for GCR intensities.
Some scientists have strongly opposed Shaviv's scenario and suggest that a GCR link to climate is at most of secondary importance to variations in CO$_2$ concentrations \citep{Royer2004GSA}. One source of difficulty in resolving this issue, which is central to understanding Earth history during the main eon of plant and animal evolution, has been uncertainties in the geological record of climate. Recent research has improved the situation in that respect. On the astronomical side, the Galaxy's spiral pattern, its rotation speed and its density variations remain uncertain.
Star formation is mainly confined to the Galaxy's spiral arms, which are lit by massive young stars. It is now generally accepted that the spiral structure seen in many galaxies is produced by density waves and probably persists for billions of years. As the Sun is a typical disk star of the Milky Way, orbiting around the Galactic centre, an important feature of the ever-changing environment experienced by the Solar System is the formation of new stars from nearby gas clouds. A large fraction of star formation in the Galaxy is accounted for by cluster formation \citep{Lada2003}, and the ages of clusters are also a guide to the changing birth rate of massive stars.
New stars that are more than about 8 solar masses (M$_\odot $) end their relatively short lives in SN explosions. These generate the shock fronts in the interstellar medium that are believed to accelerate GCR to energies in the range from 10$^{6}$ eV to 10$^{17}$ eV \citep{Berezhko2007}. The GCR primary particles are mainly protons ($\approx$ 91\%) and nuclei, and their energy density of 1 eV cm$^{-3}$ is comparable to the energy densities of the Galactic magnetic fields and the interstellar gas pressure, so that GCR play an important part in the dynamics and evolution of the interstellar medium \citep{Boulares1990AjP}.

In a model based on observed positions of the spiral arms, and the relative speed of the Sun, $\Omega_0$, with respect to the pattern speed, $\Omega_P$, of the spiral structure, $\Omega_0 - \Omega_P$, as estimated from the literature, \citet{Shaviv2002PRL} inferred the GCR flux experienced by the Solar System as it travelled through four spiral arms. The flux reached a maximum after each encounter with a spiral arm, and then went to a minimum in the dark spaces between the arms. The interval between spiral arm visits was judged to be $\approx$ 140 Myr.
Unfortunately the overall structure of the Milky Way is hard to see because of our position within the disk, and only recently has it been generally accepted that the Galaxy possesses a central bar. Although observations of the 21-cm hydrogen line show a four-arm spiral structure, there are still suggestions that the Milky Way is mainly a two-arm spiral galaxy.

In this paper the aim is to use the least model-dependent approach to the course of events in the past 500 Myr, by deriving the star formation rates and supernova rates directly from open star clusters in the solar neighbourhood, and using the SN rate as a proxy for the GCR flux to the Solar System.
The paper is organized as follows. In Sect. \ref{Sec2} the cluster formation rates over the past 500 Myr are derived directly from open star clusters in the solar neighbourhood, and then used in Sect. \ref{Sec3} for computing the local SN rates as a proxy for the GCR flux to the Solar System. In addition local features of Galactic structure are inferred from contrasting histories in star formation outside and inside the solar circle. In a effort to test the results in Sect. \ref{Sec3} a numerical model is used in \ref{Sec4} to simulate the birth, dynamics and lifetimes of open clusters, and perform exactly the same analysis as done with the observational data of open clusters. To complete the astrophysics, Sect. \ref{Sec5} shows how a close SN ($<$ 300 pc) results in a short and sharp spike in the GCR flux.

In Sect. \ref{Sec6}, where attention turns to consequences of the changing GCR flux for the Earth's climate, those spikes due to the nearest SN events are offered as an explanation of relatively sudden and short-lived falls in sea level, with brief glaciations causing the marine regressions. Comparisons of SN rates with the terrestrial climate record on longer time scales confirm the match between major periods of glaciation and persistently high GCR fluxes. Given this evidence for large effects on climate, large impacts on life are also to be expected. Section \ref{Sec7} explores an evident link between SN rates and evolutionary history, as manifest in variations in marine biodiversity. Also correlating with SN rates are variations in the Earth's carbon cycle involving bio-productivity and (in anti-correlation) carbon dioxide concentrations, during the past 500 Myr.
Sect. \ref{Sec8} discusses some implications of the paper's results, for astrophysics, palaeoclimatology and the history of life, and Sect. \ref{conc} offers brief conclusions that focus on the empirical evidence for a large impact of SNs on the prosperity of the Earth's biosphere.

\section{OPEN STAR CLUSTERS IN THE EARTH'S GALACTIC VICINITY}
\label{Sec2}
Avoiding any preconception of the precise structure of the Galaxy or of the Solar System's motion through it, the present work will reconstruct the star formation in the solar neighbourhood during the last 500 Myr from open star clusters, with a view to inferring the local SN rate as a proxy for GCR \citep{GCRSTARBURST2009}. It is generally believed that nearly all star formation has occurred in open clusters, where stars made from the same gas cloud remain for a certain time bround together by gravitation. Within each cluster the stars therefore have the same chemical composition and age. The Pleiades cluster for example, known since ancient times, is 150 pc away and estimated to be 142 Myr old, which assigns it to the Cretaceous Period on the Earth.
The justification for using the number of open clusters as a proxy for large massive stars is the evidence that the massive SN progenitor stars are mainly formed in the central region of rich stellar clusters \citep{Zinnecker1993,Hillenbrand1997,Zinnecker2007}. Secondly, of the open clusters made and embedded in giant molecular clouds, only a small fraction survive the first few million years and become visible open clusters. This small fraction of surviving open clusters are likely to have been the initially most rich clusters \citep{Lada2003}. The formation rates of open clusters are therefore used as a proxy for the formation of SNs.

\begin{figure}
\centering
\includegraphics[width=84mm]{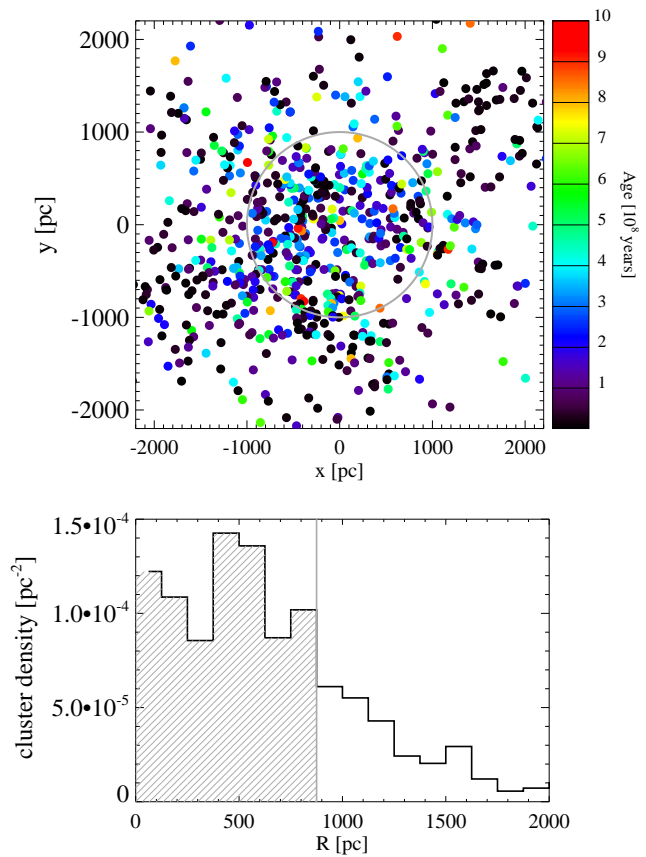}
\caption{ Top panel: The distribution of open clusters in the neighbourhood of the Solar System, plotted on the Galactic plane. The grey circle is at 1.0 kpc from the Solar System which is located at the centre of the plot. The colours denote the ages of the clusters, most of which are relatively young. Lower panel: The observed density of open clusters at increasing distances. With more and more undetected open clusters at long range, the grey line gives the maximum size for a "nearly complete" sample.}
\label{Webdadist}
\end{figure}

The WEBDA Open Cluster Database (2009) contains about 1300 open star clusters with ages between 10$^{6}$ and 10$^{10}$ years and at distances between 0.04 and 13 kiloparsec (kpc). Only a subset of these are suitable for the historical analysis of local SN rates. The clusters that emerge from the embedded phase gradually decay by the loss of stars, due to internal close encounters or to external encounters with massive clouds or other clusters. As a result the number of detectable clusters in a generation declines over time, until after $\approx$ 500 Myr most have evaporated. Figure \ref{Webdadist} (top panel) shows the distribution of the clusters within 2 kpc of the Solar System in the Galactic plane, and the colour-coded ages make it clear that the great majority of clusters are relatively young and only a few are more than 500 Myr old.

\begin{figure}
\centering
\includegraphics[width=84mm]{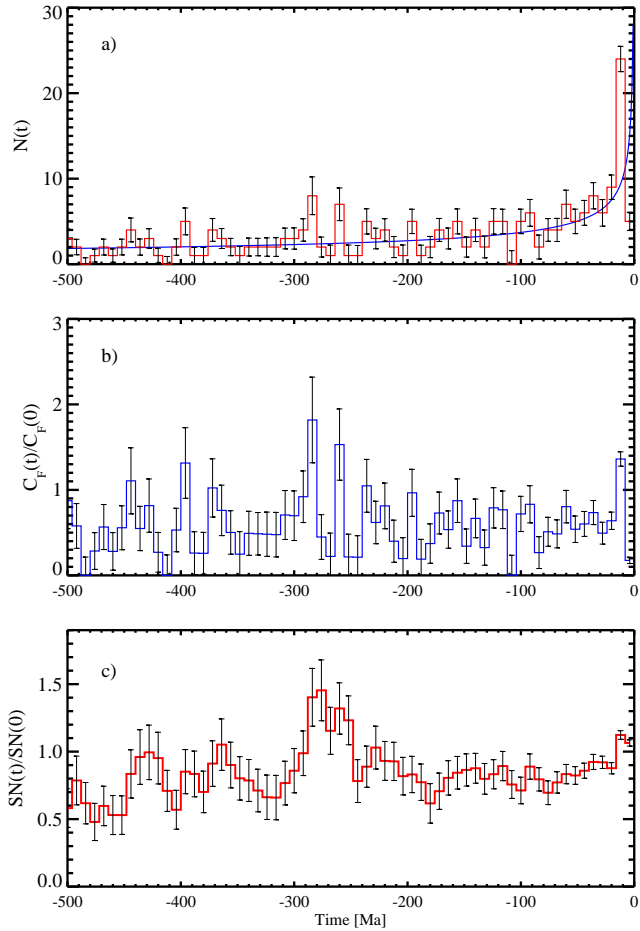}
\caption{To derive the variations in the local supernova rate within the solar neighbourhood, over 500 million years (Myr), the numbers of open star clusters within $\approx$ 0.85 kpc of the Solar System, that originated in each 8 Myr bin, are first plotted in (a). When the decay is taken into account, the formation rates of clusters over 500 Myr are derived in (b) and normalized by taking the average of the 24-16 and 16-8 Myr bins. In (c) the application of the supernova response function illustrated in Fig. \ref{SNresponsfunction_II} gives the SN rate per 8 Myr. There is a large variation between the lowest and highest rates. The calculation of error bars is explained in the text (Sect. \ref{Sec3}). Each plot starts at -510 Myr.
}\label{SN_WEBDAclusters_SOLAR}
\end{figure}

A second criterion concerns the distances of the clusters, where the problem is observational. Open clusters are hard to identify at large distances, and Fig. \ref{Webdadist} (lower panel) shows the spatial density of clusters in the WEBDA database falling away markedly beyond 1 kpc in the Galactic plane. Therefore in order to have a nearly complete statistical ensemble, selected clusters are restricted to within 0.85 kpc of the Solar System, where there are 273 clusters within 0.3 kpc above or below the Galactic plane, and with ages less 500 Myr. This sample is sufficiently large and statistically almost complete \citep{Wielen1971AA,Piskunov2006AA}. A retrospective view of changes in star formation over the long time scale of interest must nevertheless take account of the evaporation of old clusters.
The number of new clusters in some volume V of the Galaxy formed in an interval of time $(t,t+\Delta)$ can be written as
\begin{equation}\label{cluster1}
N(t,\Delta) = \int_V q(r,t) \; dV \;\Delta
\end{equation}
where $q(r,t)$ is the production of clusters per unit of time and volume, and where $\Delta$ is short compared to the lifetime of clusters emerging from the embedded phase and will be set to 8 Myr in the following. Then the surviving number of clusters of generation formed at time interval $t',t'+\Delta$ at a later time $t$ can be written as
\begin{equation}\label{cluster2}
\Psi(t-t',\Delta) = N(t',\Delta) \; \Gamma(t-t') \;
\end{equation}
where $\Gamma (t-t')$ is a function describing the decay of the number of clusters. If in a region of the Galaxy the present number of clusters as a function of age and the decay function is known, the birth rate of clusters can be determined as
\begin{equation}\label{cluster3}
N(t',\Delta) = \frac{\Psi(t-t',\Delta)}{\Gamma(t-t')}
\end{equation}
This resulting loss of old clusters can be seen in Fig. \ref{SN_WEBDAclusters_SOLAR}a, which shows the number of observed clusters within 0.85 kpc in the Galactic plane as a function of age, in intervals of 8 Myr. The blue curve represents the $\Gamma (t-t')$ fall-off with increasing age as a simple power law for the decay \citep{Chandar2006ApJ,Marcos2008ApJ,Gieles2010}, i.e.
\begin{equation}\label{dN}
\Gamma(t-t') = a (t-t')^{-\alpha}
\end{equation}
where $\alpha$=0.50$\pm$0.1 is used in Fig. \ref{SN_WEBDAclusters_SOLAR}a. It is commonly assumed that the cluster formation rate is constant in time and the decay relation $\Gamma(t)$ is determined from observations. That assumption will not be made here. Instead, what follows is based on the deviations from the decay law Eq. \ref{dN}. Using Eq. \ref{cluster3} the deviations can be considered a result of temporally (and spatially) varying cluster formation rates, as shown in Fig. \ref{SN_WEBDAclusters_SOLAR}b. The low count of clusters less than 8 Myr old, compared with that in the 8-16 Myr bin, is probably due to their concealment by natal dust clouds that have not yet dissipated. The more general fall, going back 500 Myr, is the result of cluster decay, and applying the decay law ensures that cluster formation fluctuates around a long-term mean, with little or no trend.
\begin{figure}
\centering
\includegraphics[width=84mm]{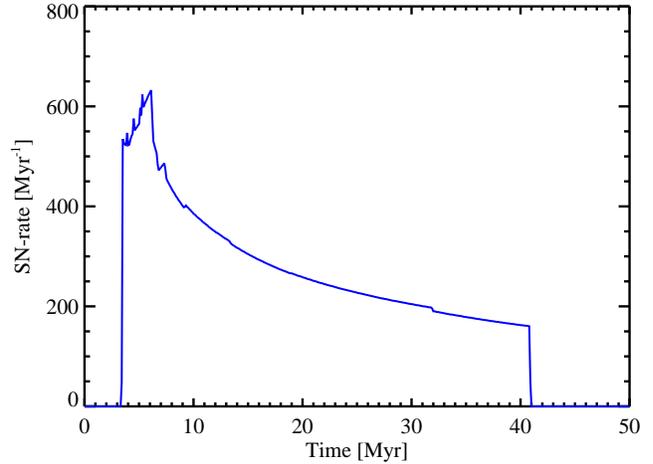}
\caption{Response function of supernovae resulting from an initial starburst of stellar mass 10$^6$ M$_\odot$ at t=0, as calculated by Starburst99 \citep{Starburst991999}. Parameters of the simulation: two initial mass function intervals, with exponents 1.3, 2.3 at mass boundaries 0.1, 0.5, 100 M$_\odot$, supernova cut-off mass 8 M$_\odot$, metallicity 0.020 and Padova track with AGB stars.}
\label{SNresponsfunction_II}
\end{figure}
\begin{figure}
\centering
\includegraphics[width=84mm]{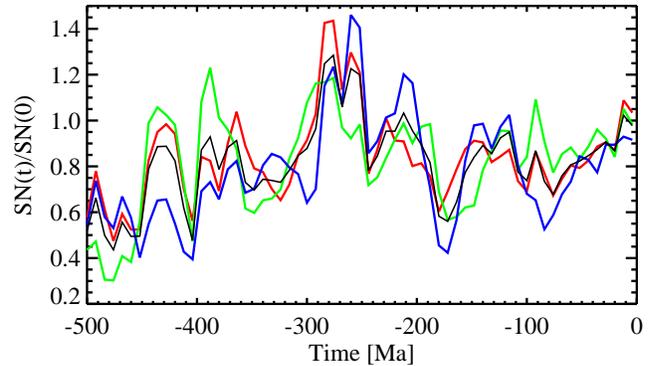}
\caption{The SN variation calculated as in Fig. \ref{SN_WEBDAclusters_SOLAR}c, but adding two other open cluster catalogues. The red curve is based on the WEBDA catalogue (273 clusters with r $\le$ 850 pc and age $\le$ 500 Myr), the green curve uses the \citet{DIAS2010} catalogue (224 clusters with distance $\le$ 850 pc and age $\le$ 500 Myr) whilst the blue curve is for the \citet{Kharchenko2005} catalogue (258 clusters with distance $\le$ 850 pc and age $\le$ 500 Myr). The black curve is an average of the red, green and blue curves, and the WEBDA results (red curve) follow it rather closely.}
\label{SN_datasets}
\end{figure}
\begin{figure}
\centering
\includegraphics[width=84mm]{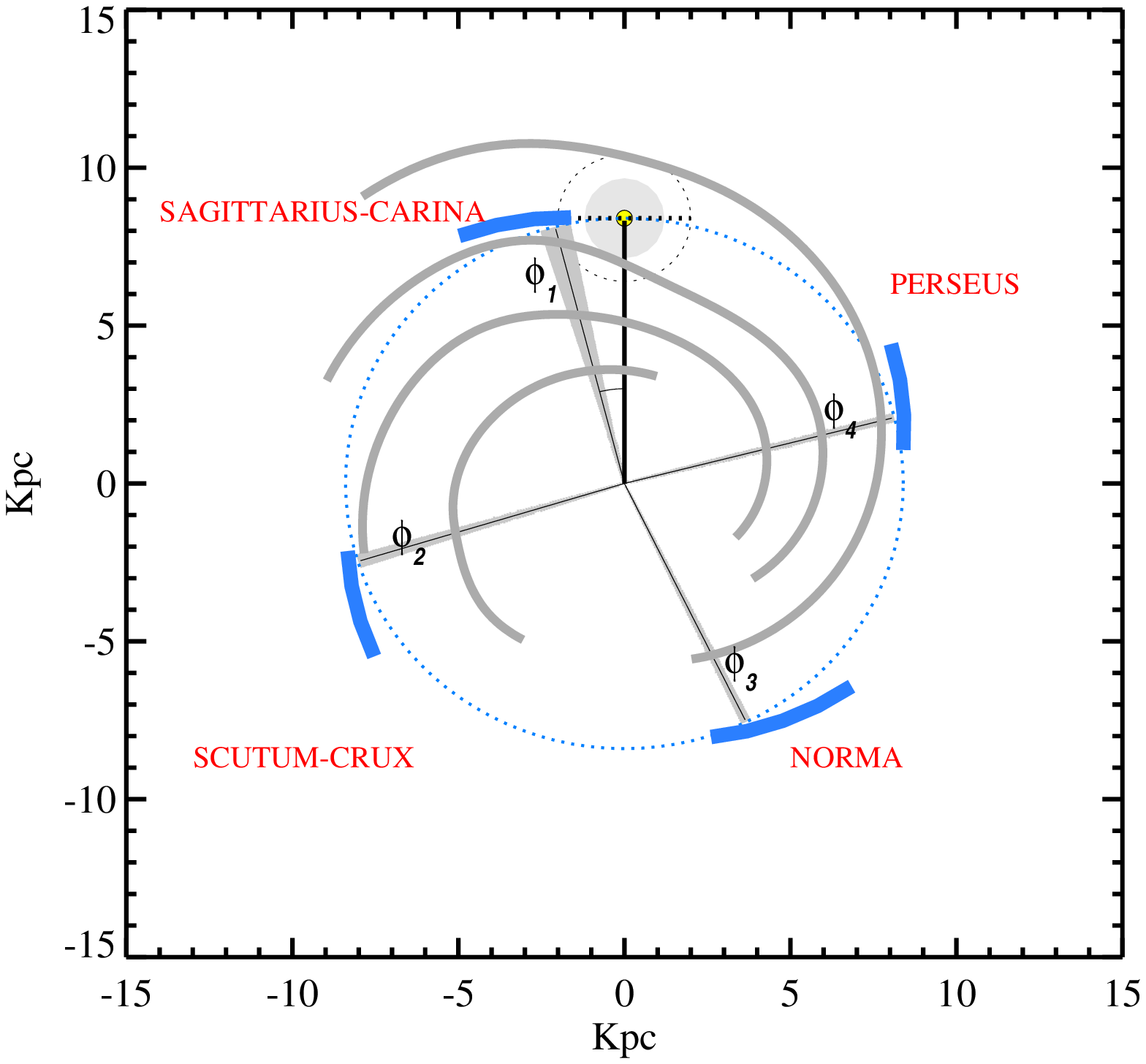}
\caption{Overview of the Milky Way. The known parts of the spiral arms are shown as the grey lines \citep{Taylor1993}. The Solar System is represented by the small yellow circle, surrounded by a grey area denoting the solar neighbourhood out to a distance of 1 kiloparsec (kpc). The two thin dotted semi-circles around the Solar System are the areas used to compare the star formation histories inside and outside the solar circle, which is shown as a blue dotted line of radius 8.5 kpc from the Galactic centre. The blue curves and the angles $\phi_i$ are the zones and positions where the Solar System encountered the maximum SN rates in front of the spiral arms (see Sect. \ref{Sec3}). The narrow grey segments show the estimated uncertainties in the maximum SN positions.}
\label{Spiral}
\end{figure}
\section{LOCAL SUPERNOVA RATES OVER 500 MILLION YEARS}
\label{Sec3}
The next step is to deduce the number of supernovae (SNs) in each time interval of 8 Myr. As open clusters are held together by their gravity, only massive groupings containing massive stars survive for long periods. The clusters form with various masses and numbers of stars, but the ratio of massive star numbers to total cluster numbers is assumed to be constant over 500 Myr.

The relative numbers of stars of various masses in a cluster is to a good approximation given by Salpeter's initial mass function (IMF) power law \citep{Salpeter1955ApJ}. In consequence the number of stars going supernova will be, on average, proportional to the number of clusters in a bin, but the occurrences of the SNs will spread into later bins. The evolution of stars in a cluster to their detonations as supernovae is simulated numerically by the Space Telescope Science Institute's Starburst99 program \citep{Starburst991999}.

As seen in Fig. \ref{SNresponsfunction_II}, if a stellar mass of 10$^6$ solar masses comes into being in an instantaneous starburst, supernovae first occur after $\approx$ 3 Myr and they continue for $\approx$ 30-40 Myr, until the last of the massive stars abruptly disappear. This form of the SN response function can be used to obtain the temporal variation in the SN rate caused by changes in the number of new clusters created in the time interval $(t,t+\Delta)$ as
\begin{equation}\label{SN}
SN(t,\Delta) = c_{sn} \int_{-\infty}^t { N(t',\Delta) R_{SN}(t-t') } \;
dt',
\end{equation}
where $c_{sn}$ is a constant, $N(t',\Delta)$ is the cluster formation history given by Eq. \ref{cluster3} and shown in Fig. \ref{SN_WEBDAclusters_SOLAR}b, and finally $R_{SN}$ is the SN response function to a starburst. A simplifying assumption is that the stars in the $N(t,\Delta)$ clusters are formed instantly at $t'$. The numerical integration is displayed in Fig. \ref{SN_WEBDAclusters_SOLAR}c which shows the SN rates as a function of time.
A spatial scale of $\approx$1 kpc and temporal time steps of 8 Myr ensure that the GCR flux has had time to equilibrate with the newly appearing sources in the region. The diffusion constant of a 1 GeV GCR particle is 0.13 kpc$^{2}$Myr$^{-1}$ (see Sect. \ref{Sec5}), which takes about 8 Myr to equilibrate over a 1 kpc region. The SN rate is normalized to the present SN rate in the solar neighbourhood by taking the average of the two 24-16 and 16-8 Myr bins, ignoring the 8-0 Myr bin rate which is misleadingly low because many new clusters are still hidden in dust. The present SN rate in the solar neighbourhood has been estimated in the range of 20-30 SNs Myr$^{-1}$kpc$^{-2}$ \citep{Grenier2000AA}, which gives $\approx$ 500-750 SNs for a typical 8 Myr bin and an area of $\pi$ kpc$^2$ (solar neighbourhood).

The resulting SN rates shown in Fig. \ref{SN_WEBDAclusters_SOLAR}c should therefore indicate the changes in GCR flux experienced by the Earth's environment due to visits to regions of the Galaxy with high or low rates of open cluster formation. The delays between formation and detonation of massive stars have the effect of partly smoothing the very large variations from bin to bin seen in cluster numbers (Fig. \ref{SN_WEBDAclusters_SOLAR}b). Nevertheless, taking the SN rate in the solar neighbourhood as a proxy for GCR at the Earth, Fig. \ref{SN_WEBDAclusters_SOLAR}c implies that persistent ionization in the Earth's atmosphere due to GCR went up and down by a factor of 2 during the last 500 Myr.
\begin{figure}
\centering
\includegraphics[width=84mm]{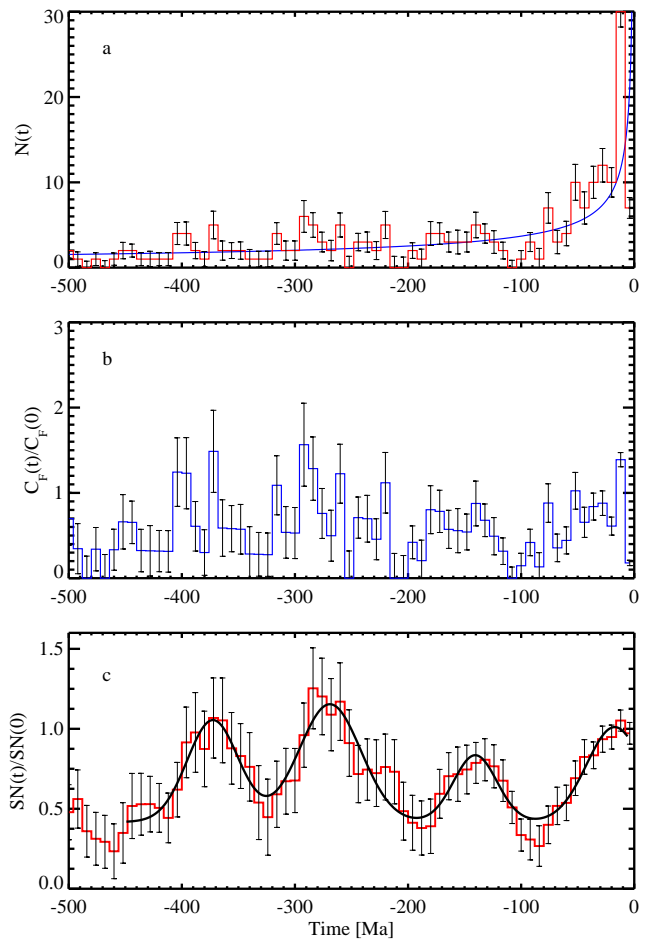}
\caption{ Variations in the local supernova rate outside the solar circle over 500 million years
(Myr). The numbers of open star clusters within 2 kpc of the Solar System, that originated in each
8 Myr bin, are plotted in (a). When the decay of clusters is taken into account, the formation rates of
clusters over 500 Myr are derived, as shown in (b), with the rate normalized by taking the average of the two 24-16 and 16-8 Myr bins. In (c), application of the supernova response function illustrated in Fig. \ref{SNresponsfunction_II} gives the supernova rate per 8 Myr. The black curve is a least square fit of Eq. \ref{M} to the data. The calculation of error bars is explained in the text. Notice the wave pattern in (c) suggesting the presence of four spiral arms, although plainly unequal. Plot starts at -510 Myr.}
\label{WEBDAclustersFIT}
\end{figure}
\begin{figure}
\centering
\includegraphics[width=84mm]{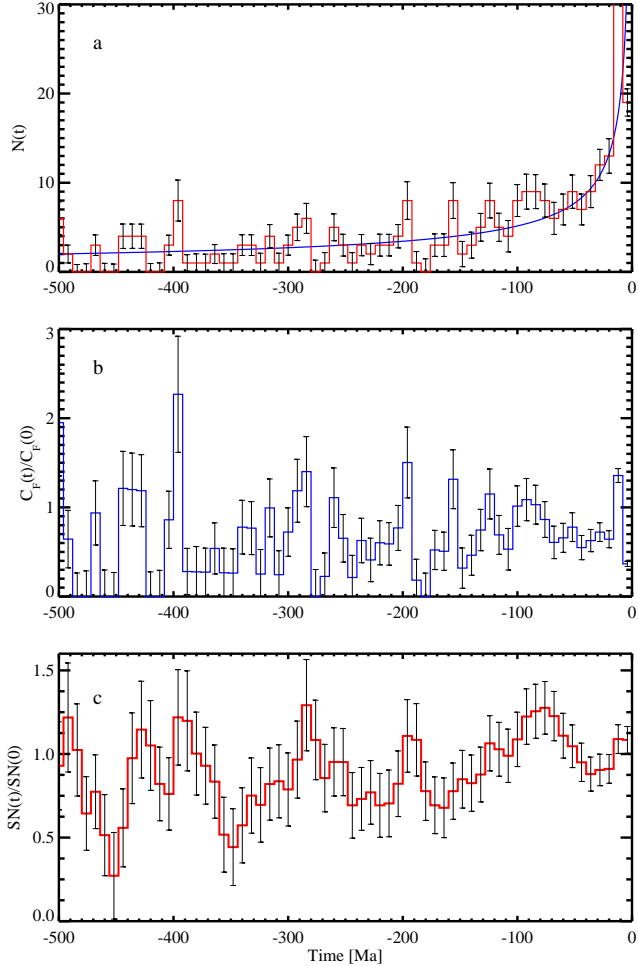}
\caption{As Fig. \ref{WEBDAclustersFIT}, but for open clusters inside the solar circle, within 2 kpc of the Solar System: (a) is the number of clusters, (b) the formation rates of clusters, and (c) the supernova rate per 8 Myr. Notice that in this case the orderly wave seen in Fig. \ref{WEBDAclustersFIT}c is replaced by a much more complex pattern. Plot starts at -510 Myr.}
\label{WEBDAclustersINDSIDE}
\end{figure}
As for the error bars in Fig. \ref{SN_WEBDAclusters_SOLAR}, the typical error in the age of a cluster is of the order 10-20\% \citep{Marcos2004NewA}. To estimate the resulting uncertainty by a bootstrap Monte Carlo method, 37\% of the cluster ages, chosen at random in a sample, are replaced by a new age which is drawn from a random normal distribution with a 20\% variance in the age and centred around the measured age. This process is repeated for 10$^3$ samples. The resulting variance in the number of clusters for each age bin is estimated and plotted as the error bars in Fig. \ref{SN_WEBDAclusters_SOLAR}a. Similarly the error bars in Figs. \ref{SN_WEBDAclusters_SOLAR}b and \ref{SN_WEBDAclusters_SOLAR}c are calculated by generating pseudo cluster distributions and calculating the resulting pseudo cluster formation rates and SN rates. Here the error bars increase with age due to the smaller relative number of observations of older clusters. A bootstrap Monte Carlo simulation adding Poisson noise on the number of clusters increases the standard variation by $\approx$ 15\% in Fig. \ref{SN_WEBDAclusters_SOLAR}c.

Evidence from isotopes made by GCR hitting meteorites while they orbited in space provides a test of whether the GCR variations in the past derived as in Fig. \ref{SN_WEBDAclusters_SOLAR} are realistic. \citet{Lavielle1999} conclude from the production rates of $^{36}$Cl in a calibration data set of 13 meteorites that the flux of cosmic rays in the Solar System during the past 10 Myr was 28\% higher than the average over the past 500 Myr. For comparison, for the SN rates derived from the open clusters in Fig. \ref{SN_WEBDAclusters_SOLAR}c the average of the two early bins -24 to -8 Myr is 32\% higher than the average over the 500 Myr - in satisfactory agreement.

As mentioned above the variation in SN rates was calculated using the WEBDA database. There are however other compilations of open clusters with differences in selection criteria which in some cases give slightly different parameters. It is therefore prudent to compare the temporal variation of SN intensity shown in Fig. \ref{SN_WEBDAclusters_SOLAR}c with inferences from other open cluster compilations to test the consistency of the results. Figure \ref{SN_datasets} show the WEBDA result (red curve) together with the widely used \citet{Dias2002,DIAS2010} catalogue (green curve) and the \citet{Kharchenko2005} catalogue (blue curve). Although there are differences, the main features are similar and the average of the three data sets (black curve) follows the WEBDA results closely. The WEBDA catalogue will be used exclusively in the remainder of the paper.

\subsection{INTERPRETING THE MILKY WAY'S STRUCTURE}
Having started with no assumptions about the configuration of the Milky Way's spiral arms, the method of analysis adopted here can infer the local Galactic structure from cluster ages and the associated supernova (SN) rates. There is ample evidence that the Galaxy is a spiral galaxy, and based on velocity-longitude maps a 4-armed spiral is detected \citep{Blitz1983,Dame2001ApJ,Vallee2008AJ} outside the solar circle extending out to about $2R_0 \approx 17$ kpc. However, inside the solar circle it is more difficult to make unambiguous statements on the spiral structure, although it seems observationally secure that the Galaxy is a barred spiral with a bar pattern speed in the range 50 - 60 $\textrm{km s}^{-1}\textrm{kpc}^{-1}$ . But the pattern speed of the spiral arms is one of the least constrained parameters and has been estimated within a range of 10 to 30 $\textrm{km s}^{-1}\textrm{kpc}^{-1}$ depending on the method used. For a recent review of the spiral pattern speeds, see \citet{Shaviv2003NewA}. One reason for the large variability in the deduced pattern speeds might be that there exist more that one pattern speed. For example \citet{Noaz2007NewA} found two pattern speed solutions to the Carina arm by tracking the birthplaces of open clusters. If the spiral pattern as seen outside the solar circle is a 4-armed density wave extending to about 17 kpc, then the dynamical stability given by the Lindblad resonance places the 4 arms from about the solar circle with a pattern speed less than $\approx$ 20 $\textrm{km s}^{-1}\textrm{kpc}^{-1}$. (See for example \citet{Shaviv2003NewA}).

The history of SN rates in Fig. \ref{SN_WEBDAclusters_SOLAR}c does not show a clear regular pattern that one might expect if the Solar System simply passed through four similar spiral arms as suggested by \citet{Shaviv2002PRL}. To verify that the present results on SN rates could none the less be compatible with a realistic Galactic structure, it will be shown that the complications in Fig. \ref{SN_WEBDAclusters_SOLAR}c seems to arise from differences in the Galactic structure inside and outside the solar circle. This is done by extending the range of open clusters from the WEBDA database to 2 kpc in order to have adequate numbers of clusters to repeat the procedure used to generate Fig. \ref{SN_WEBDAclusters_SOLAR} but now differentiating between clusters lying outside or inside the solar circle. Figure \ref{Spiral} shows the solar circle in relation to observed spiral arms, and around the position of the Solar System are two semicircles of 2 kpc radius where the clusters are to be incorporated.

Starting with the outside semicircle, one gets the results shown in Fig. \ref{WEBDAclustersFIT}. The four maxima in Fig. \ref{WEBDAclustersFIT}c can be interpreted as the Solar System's encounters with four spiral arms, and used to extract information about the spiral structure just outside the solar circle. The model used to fit the data, with the black curve in Fig. \ref{WEBDAclustersFIT}c, is given by
\begin{equation}\label{M}
M =A_0+\sum_{i=1}^4 A_i
\exp [-(t-t^i_0)^2/2\sigma_i^2],
\end{equation}
which consists of four Gaussian functions. The parameters of the model are determined by a least square minimization and are shown in Table 1. These results suggest that maxima in SN occurred at time intervals $\Delta T$ =
(123.3$\pm$12.9, 128.5$\pm$5.8, 103.5$\pm$4.9) Myr, averaging to $<\Delta T>$ =
118.4$\pm$7.8 Myr. The positions of the maxima are also shown in Fig. \ref{Spiral}. With the spiral pattern rotating at a smaller angular frequency $\Omega_P$, the Solar System moves in and out of the spiral arms with the relative angular frequency
\begin{equation}\label{Omega}
\Delta\Omega = \Omega_0 - \Omega_P
\end{equation}
where $\Omega_0 = 30.3 \pm 0.9 \textrm{km s}^{-1}\textrm{kpc}^{-1}$ is the Solar System rotation frequency \citep{ROTATIONCURVE2009ApJ}, giving the Solar System's rotation frequency relative to the spiral structure as
\begin{equation}\label{OmegaP}
\Omega_0 - \Omega_P = \frac{\pi}{2<\Delta T>}\frac{10^3}{1.023} =  13.0 \pm 0.9 ~ \textrm{km s}^{-1} \textrm{kpc}^{-1}
\end{equation}
where $<\Delta T>$ is in units of Myr. The relative pattern speed is therefore $\Omega_P/\Omega_0 = 0.57 \pm 0.5$, or $\Omega_P = 17.3 \pm 1.8 ~ \textrm{km s}^{-1} \textrm{kpc}^{-1}$.

Although there has been a large range of estimates of $\Omega_P$, the value found here is in agreement a range of values found in the literature\citep{Shaviv2003NewA} of $\Omega_P \approx 16 - 20 ~ \textrm{km s}^{-1} \textrm{kpc}^{-1}$.
Further it is consistent with four spiral arm encounters of the Solar System over the last 500 Myr as seen in geological records \citep{Shaviv2003NewA,Gies2005ApJ,Svensmark2006AN}.

Inside the solar circle, as shown in Fig. \ref{WEBDAclustersINDSIDE}, the pattern was not clearly periodic, and SN rates on that side plainly contributed to the irregularities seen in Fig. \ref{SN_WEBDAclusters_SOLAR}c. A possible explanation for the differences, outside and inside, comes when considering the dynamical stability of the Galaxy. Using a recently estimated Galactic rotation curve \citep{ROTATIONCURVE2009ApJ} and the pattern speed $\Omega_P$ determined above, one finds that the Solar System is right on the inner edge of the four-arm stability region (bounded by the inner 1:4 Lindblad resonance). The observed difference between Figs. \ref{WEBDAclustersFIT} and \ref{WEBDAclustersINDSIDE} may therefore be an unsurprising result of the four-arm structure losing its stability inside the solar circle. It is however not conclusive, and there are other suggestions about the position relative to the Solar System of the Lindblad resonances \citep{Lepine2011MNRAS}.

The variations in SN rates outside the solar circle, with maxima approximately every 120 Myr, seem more likely to conform to well-known spiral arms of the Milky Way. The maxima in SN rates seen in Fig. \ref{WEBDAclustersFIT}c, applied to the overview of the Galaxy in Fig. \ref{Spiral}, determine the angles shown there for maximum SN activity. They match well with the known positions of spiral arms, when one notes that the maximal SN regions are right in front of the arms, in agreement with an expected average delay of the SN explosions of about 18 Myr after cluster formation. Finally, the sizes of the maxima in Fig. \ref{WEBDAclustersFIT}c indicate that the star formation in the Scutum-Crux arm during the Solar System's passage 140 Myr ago was considerably weaker than in the other arms.
\begin{table}
\caption{Model parameters of the function used in Fig. \ref{WEBDAclustersFIT}c (black curve) are related to passages through the Sagittarius-Carina, Perseus, Norma and Scutum-Crux arms. A$_0 =$0.4$\pm$0.0.}
\begin{tabular}{c c c c c} \hline\hline
Spiral arm & Perseus & Norma & Scutum-Crux & Sgr-Car \\\hline
$t_\textrm{max}$ (Myr) & -373.4 $\pm$ 2.2 & -270.4 $\pm$ 2.4 & -140.6 $\pm$ 3.5 & -18.9 $\pm$ 7.6\\
$\sigma$ (Myr) & 21.7 $\pm$ 2.7 & 25.3 $\pm$ 3.1 & 21.0 $\pm$ 4.6 & 23.4 $\pm$ 7.0\\
A & 0.7 $\pm$ 0.1 & 0.7 $\pm$ 0.1 & 0.4 $\pm$ 0.1 & 0.6 $\pm$ 0.1\\
$\phi$ (deg.) & 284.4 $\pm$ 1.9 & 205.9 $\pm$ 2.0 & 107.1 $\pm$ 3.0 & 14.4 $\pm$ 6.5\\\hline
\end{tabular}
\end{table}
\section{SIMULATING CLUSTER HISTORIES}
\label{Sec4}
Is it really possible to extract past star formation rates and SN rates from the age distribution of open clusters in the solar neighbourhood? As open clusters are born from their parent gas clouds with small random velocity dispersions there will be a dispersion of the positions of open clusters as a function of time. Moreover clusters in near circular orbits at larger Galactic radii are overtaken by clusters at smaller radii, which results in a shearing of an initial area of the Galactic plane in the $\phi$ direction as a function of time (also called phase mixing). To examine this question, a model will simulate the birth, dynamics and lifetimes of open clusters numerically, and perform exactly the same analysis as done above for the observational data of open clusters.

The gravitational potential of the Galaxy will not include the gravitational effects of spiral arms which, for the relatively short time period of 500 Myr, are believed to be of less importance. The potential is therefore approximated by an axisymmetrical potential as \citep{Faucher2006ApJ}
\begin{equation}\label{GpotentialI}
\Phi_G (R,z) = \Phi_{dh} (R,z)+\Phi_b(R)+\Phi_n(R)
\end{equation}
where the disk halo potential is
\begin{equation}\label{GpotentialII}
\Phi_{dh} (R,z)= \frac{-GM_{dh}}{\sqrt{(a_G+\sum_{i=1}^3 \beta_i\sqrt{z^2+h_i^2})^2+b_{dh}^2 + R^2 }}
\end{equation}
and the potential for the bulge and nucleus is
\begin{equation}\label{GpotentialIII}
\Phi_{b,n} (R,z)= \frac{-GM_{b,n}}{\sqrt{b_{n,b}^2+ R^2 }}
\end{equation}
where $R$ is the radial distance from the Galactic centre and $z$ is the height perpendicular to the Galactic plane. The constants in the gravitational potential are given in Table 2. The equations of motion in cylindrical coordinates (see for example \citet{Binney1987}) are
\begin{eqnarray}\label{EqmotionI}
\ddot{R} &=& -\frac{\partial \Phi_G (R,z)}{\partial R} + \frac{L^2_z}{R^3} \\\nonumber
\ddot{z} &=&- \frac{\partial \Phi_G (R,z)}{\partial z} \\\nonumber
L_z &=& R^2 \dot{\phi}
\end{eqnarray}
where $L_z$ is the conserved angular momentum around the $z$-axis.
With this simple set of equations it is now possible to numerically simulate the number of clusters and their ages in the solar neighbourhood as a function of time, taking account of the dispersion of clusters and loss by evaporation.
\subsection{Velocity dispersion of open clusters}
\begin{figure}
\centering
\includegraphics[width=84mm]{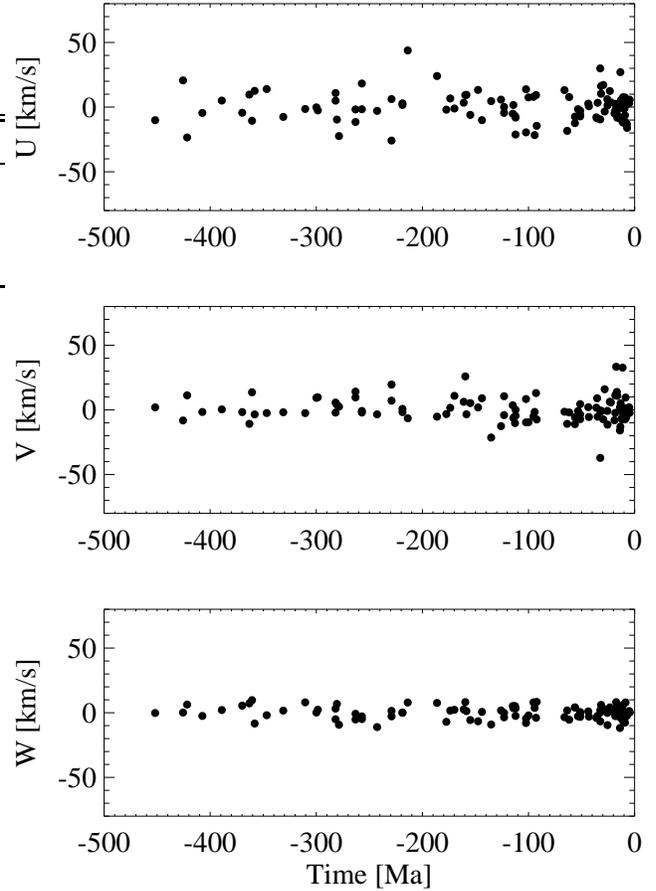}
\caption{\label{V_distribution_age}
The three components of velocities $(U,V,W)$ for 105 open clusters within 850 pc of the Solar System and ages less than 500 Myr are shown as a function of open cluster age. The three panels are from top to bottom the $U$, $V$ and $W$ components of the velocity. There is no noticeable increase in the variance over the 500 Myr period.}
\end{figure}
Open clusters for which both proper motions and mean radial velocities are specified in the WEBDA database are analysed in order to estimate the velocity dispersion. Restricting the data to open clusters within a radius of 850 pc of the Solar System and with ages less than 500 Myr reduces the number to 105 open clusters. Their measured proper motions and mean radial velocities include the following components: 1) $\textbf{v}_i$, the velocity of the cluster, 2) $\textbf{v}_\odot$, the velocity of the Solar System, 3) $\textbf{v}_{rot}$, a velocity component due to Galactic rotation, and 4) a measurement error.

The velocities are initially transformed to (U,V,W) in the local system of rest (LSR), where U is in the direction of the anti-Galactic center, V in the direction of Galactic rotation, and W in the direction of Galactic north. Applying corrections for the Solar System's velocity with respect to the LSR, $(U_{\odot}, V_\odot, W_\odot)$ = (-9, 12, 7) km/s, and for Galactic rotation,
\begin{equation}\label{Velocities}
\textbf{v}_i = \textbf{V}_i - \textbf{v}_\odot - \textbf{v}_{rot}
\end{equation}
where the index $i$ refers to each of the 105 open clusters. The velocities $v_i$ display a systematic variation as a function of Galactic latitude $l$, which is removed. The resulting $(U,V,W)$ velocities are shown as a function of age in the three panels of Fig. \ref{V_distribution_age}. A reasonable assumption is that the components of velocities $(U,V,W)$ have a Gaussian distribution and therefore the distribution of the velocities squared will be the gamma distribution $\gamma(\frac{1}{2},\frac{1}{2})$. Arranging the data into bins of size 10 km$^2/$s$^2$ and fitting the resulting distribution for each of the velocity components $(U,V,W)$ gives the variance of each velocity component $\sigma$. However all velocities contain a small measurement error which adds to the true velocity dispersions. If the measurement error is also assumed to have a Gaussian distribution, the velocity can be written
\begin{equation}
\textbf{v}_i = \widetilde{\textbf{v}_i} + \textbf{e}_i
\end{equation}
where $\textbf{v}_i$ is the required velocity and $\textbf{e}_i$ is the small error. The velocity dispersion becomes
\begin{eqnarray}
\sigma^2_U &=& \widetilde{\sigma}^2_U +\epsilon^2_{U} \\\nonumber
\sigma^2_V &=& \widetilde{\sigma}^2_V +\epsilon^2_{V} \\\nonumber
\sigma^2_W &=& \widetilde{\sigma}^2_W +\epsilon^2_{W}
\end{eqnarray}
where $\sigma$ on the left hand side is the estimated velocity dispersion, $\widetilde{\sigma}$ is the required variance and $\epsilon$ is the error in estimating the velocity.

The final step is to perform a bootstrapping Monte-Carlo simulation where $e^{-1} \approx$ 37 \% of the velocities $\overrightarrow{v_i}$ are chosen at random, and a small Gaussian distributed velocity component $\varepsilon$ is added with $\sigma^2_\varepsilon$ = 1 km$^2$/s$^2$. For each of 10$^3$ realizations one determines the velocity dispersions and so probes the sensitivity of the parameters. The average dispersions of the ensemble are found to be
\begin{eqnarray}
\label{Vdispersions}
\sigma_U &=& 5.7 \pm 1.4 ~~\textrm{km/s}\\\nonumber
\sigma_V &=& 3.2 \pm 0.8 ~~\textrm{km/s}\\\nonumber
\sigma_W &=& 3.2 \pm 0.4 ~~\textrm{km/s}
\end{eqnarray}
These values are similar to those found for young clusters by \citet{Piskunov2006AA} and are for ages $\approx$ 6 Myr, $(\sigma_{\textrm{u}},\sigma_\textrm{v},\sigma_\textrm{w}) = (7.2,\;4.1,\;2.6)\; \textrm{km/s}$ but with the important distinction that the dispersions are found not to increase over the 500 Myr time span used in this work, as is also indicated by visual inspection of Fig. \ref{V_distribution_age}. Although the velocity dispersions in Eq. \ref{Vdispersions} still contain an unknown small added error, the foregoing analysis gives credence to the use made of open cluster data over the past 500 Myr.

\subsection{Numerical procedure}
\begin{figure}
\centering
\includegraphics[width=84mm]{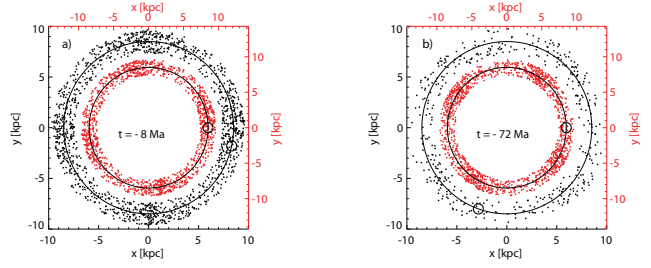}
\caption{Snapshot illustrations of a model of the dynamics of open star clusters that tests the effect of cluster dispersion on the reconstruction of SN rates in the past, as was done with real data in Sect. \ref{Sec3}. The model tracks the motions of clusters following their formation within a 2 kpc annulus in the Galactic plane containing the solar circle, which has a radius of 8.5 kpc. The red coordinate axes are scaled smaller than the black axes by a factor of 0.7 for better viewing of initial and final positions of clusters. Plotted using the red axes, the red points show the initial positions of clusters born at times a) -8 Myr and b) -72 Myr, assuming a 4-armed spiral structure with the arms separated by 90 deg. in phase angle. The small circle with radius 0.5 kpc indicates the solar neighbourhood. Plotted using the black axes, the black points show the positions of the surviving fraction of clusters at time $t$=0 Myr, after integrating the trajectories for $t$=8 Myr and $t$=72 Myr, respectively. The position of the solar neighbourhood, again shown by the small circle, rotates more than 90 deg over 72 Myr. Note that the number of clusters has visibly diminished after 72 Myr, by the model's allowance for a randomized loss of clusters (see text).
\label{phaseplotarms4plot}}
\end{figure}
\begin{figure}
\centering
\includegraphics[width=84mm]{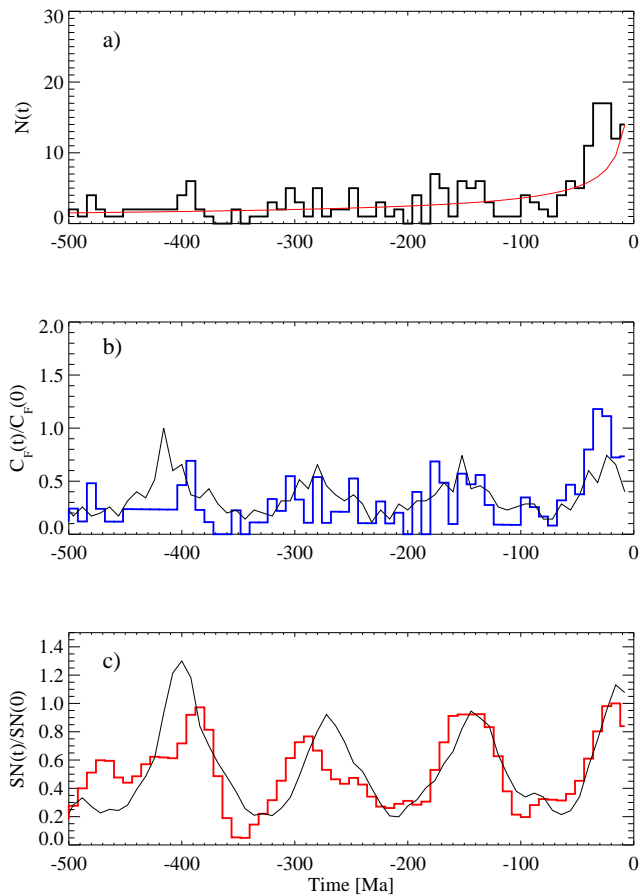}
\caption{
Example of a simulated reconstruction of the SN rate based on the age distribution of open clusters over 500 million years (Myr) within the solar neighbourhood (a). When the decay is taken into account, the formation rates of clusters over 500 Myr are derived in the blue curve in (b), where the black curve is what the rates in the solar neighbourhood should have been according to the initial input into the simulation. In (c) the application of the supernova response function illustrated in Fig. \ref{SNresponsfunction_II} gives the SN rate per 8 Myr (red curve), whilst the black curve is the SN reconstruction based directly on the input number of clusters, as in the black curve in (b). It is a general feature of the simulations that the oldest parts give less accurate reconstructions.
\label{SIMExample}}
\end{figure}
\begin{figure}
\centering
\includegraphics[width=84mm]{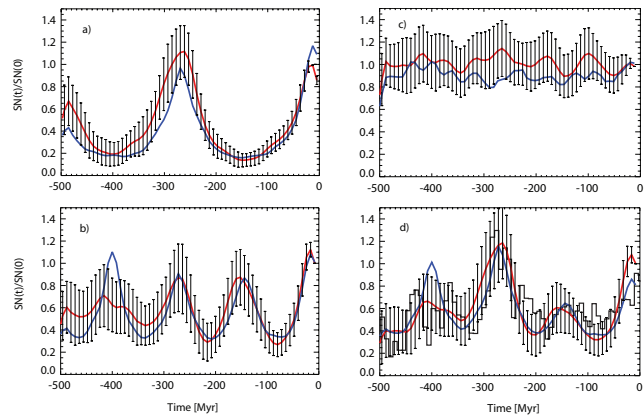}
\caption{
Results from the model of cluster dynamics applied over 500 Myr give normalized SN rates near the Solar System. The blue curves show the rates that would be inferred directly from the modelled clusters in the solar neighbourhood, had they been observable at the time of their formation. To show results obtained by looking back in time from the present age distribution of nearby clusters, each of the red curves is inferred from 100 realizations of the simulated age distribution at $t=0$, which take account of cluster dynamics. The error bars show the 1-$\sigma$ variance of the realizations. In Panel a) the simulated formation of clusters implied in the blue curves accords with a spiral structure with two equal arms 180 deg. apart and with velocity dispersions given in the text. Panel b) is similar to a) but for a spiral structure with four equal arms 90 deg. apart, whilst in Panel c) the cluster formation is spatially and temporally homogeneous and the reconstructed SN rate is approximately constant. Finally, Panel d) simulates the results from real data in Sect. \ref{Sec3}, shown in the black histogram, by using unequal spiral arm amplitudes $p_{i} =(1.0,1.5,0.7,0.8)$ in generating the birth of clusters, and with velocity dispersions given in the text. Notice that a fairly good agreement persists between the red and blue curves, meaning that it is possible to reconstruct the SN variation from the age distribution in the solar neighbourhood, although the uncertainty is larger for the oldest part.
\label{SIM4FIG}}
\end{figure}
The aim is to simulate the formation and dynamics of open clusters in the Galaxy, and to be able to modulate the effect of spatially and temporally variations in cluster formation by the presence of for example spiral arms. The simulation is confined to an annulus with inner radius $R_{\textrm{min}} = 7.4$ kpc and outer radius $R_{\textrm{max}} = 9.6$ kpc, with the solar circle at $R_{\odot} = 8.5$ kpc. The effect of the spiral arms is simulated by a density variation in the formation of new clusters, using the Gaussian relation
\begin{equation}
\label{P} P(\phi,t)= p_{0} +\sum_{i=1}^{4}p_{i} \exp{
\left [ \frac{(\phi - \phi_{i}(t))^2}{2\sigma_{i}^2}\right] }. \nonumber
\end{equation}
where $\phi$ is the angular coordinate, $\phi_{i}(t)$ is the angular position of the $i$'th arm at time $t$, $p_{i}$ is the amplitude of the density maxima, $p_{0}$ is a constant, and $\sigma_{i}$ is the width of the density maxima. Finally $P(\phi,t)$ is normalized so its maximum is one, with imposed periodic boundary conditions in $\phi$. The temporal dependence of $\phi_{i}(t)$ makes it possible to simulate a pattern speed for the spiral structure. The pattern speed $\Omega_P$ is chosen so that $\Delta\Omega$ of Eq. \ref{OmegaP} corresponds to $\Delta T =$ 128 Myr between spiral arm passages of the Solar System.
The numerical procedure begins by simulating 10$^4$ trajectories originating at random positions distributed uniformly in the $(R,\phi)$ plane, in the annulus of 2 kpc containing the solar circle, and exponentially distributed in the $z$ direction with scale height of 50 pc. Integration of the equations of motion over a time period $T_n$, where $T_n = [8, 16, 24, \ldots , 512]$ Myr, then gives 10$^4$ new trajectories for each $T_n$.

This basic set of trajectories is then modulated using Eq. \ref{P} to simulate either a homogeneous stationary pattern or a dynamical spatial structure caused by a 2-armed or a 4-armed spiral pattern whereby clusters are born with a higher probability within a spiral arm than between spiral arms, and are rotating with a spiral pattern speed $\Omega_P$. In addition cluster lifetimes are simulated by removing a fraction (chosen at random) of cluster trajectories of age $T_n$, in agreement with the decay law in Eq. \ref{dN} for the evaporation or disintegration of clusters.

Figure \ref{phaseplotarms4plot} is an example which displays a subset of the simulated cluster histories. The red annuli in panels a) and b) show the initial positions in the Galactic plane of newly formed clusters at times -8 and -72 Myr respectively, with the imposed spatial modulation around the annulus corresponding to the 4-armed spiral structure. The pattern is moving, so that there is a 128 Myr period between encounters of the solar neighbourhood (indicated by the small black circle) and the highest rates of cluster formation found in the arms. Using a larger scale (black coordinate axes) for better viewing, the black annuli of points in Fig. \ref{phaseplotarms4plot} show the positions of clusters integrated from $t =$ -8 and -72 Myr to $t=0$. Notice that for clusters of age 72 Myr the number surviving is already visibly reduced.

Figure \ref{SIMExample}a shows one simulated realization of the age distribution over the 500 Myr period. The cluster formation rate is chosen so that the number of clusters in the solar neighbourhood at $t=0$ corresponds realistically to what is observed. Correcting for the decay of clusters leads to the cluster formation rate over the 500 Myr as shown in the blue curve in Fig. \ref{SIMExample}b. The black curve in that panel is the modelled input of the number of clusters in the solar neighbourhood at each instant of time. Finally Fig. \ref{SIMExample}c is the derived SN variation (red curve), together with the SN variation based directly on the input clusters in the solar neighbourhood (black curve). One sees a fairly good agreement over the period although the oldest part is less satisfactory.

Figure \ref{SIM4FIG} shows applications of the cluster dispersion model to various star formation histories over 500 Myr, in a 1 kpc region around the Solar System, assuming different models of the Galaxy. In each panel the blue curve shows the "true" history of SN rates in the model, based on a simulated formation of open clusters that would have been observable by astronomers, had they been alive all those millions of years ago. Each red curve shows the history of SN rates "inferred" from the modelled age distribution of surviving clusters in the solar neighbourhood at $t=0$, using the method applied to real observations in Sect. \ref{Sec3}. Fig. \ref{SIM4FIG}a assumes a 2-armed spiral structure with equal sized arms separated by 180 deg., $p_{i} =(1,0,1,0)$ and $p_0=0$ and velocity dispersions $\sigma_{\textrm{u}},\sigma_\textrm{v},\sigma_\textrm{w}) = (4.3, 2.4, 2.8)$ given in Eq. \ref{Vdispersions}. The "inferred" SN history in the red curve is averaged over 100 realizations of the simulation. There is overall agreement with the "true" history in the the blue curve, although it is less good in the earliest times ($<$ -400 Myr). Fig. \ref{SIM4FIG}b is similar, but for a 4-armed spiral structure with equal sized arms separated by 90 deg., $p_{i} =(1,1,1,1)$ and $p_0=0$ and velocity dispersions $\sigma_{\textrm{u}},\sigma_\textrm{v},\sigma_\textrm{w}) = (4.3, 2.4, 2.8)$. The "inferred" SN history in the red curve is averaged over 100 realizations of the simulation. Again there is overall agreement with the "true" history in the the blue curve, less good in the earliest times ($<$ -400 Myr). A situation with spatially and temporally homogeneous cluster formation is shown in Fig. \ref{SIM4FIG}c the SN intensity over the 500 Myr period displays a fairly constant SN rate with only small fluctuations.

Finally the real observations are simulated in Fig. \ref{SIM4FIG}d by using unequal spiral arms as specified in the caption and with velocity dispersions $(\sigma_{\textrm{u}},\sigma_\textrm{v},\sigma_\textrm{w})$. The red curve is the reconstructed SN variation from 100 realizations of the simulation, with error bars showing the 1-$\sigma$ variance of the realizations. The black histogram is the reconstruction based on real observations as seen in Fig. \ref{SN_WEBDAclusters_SOLAR}c. The correspondence between the curves is satisfactory, although it falters again at about -400 Myr. Comments on the performance of the simulations are deferred to the Discussion in Sect. \ref{Sec8}.
\begin{table}
\caption{Parameters of the model Galactic potential Eqs. \ref{GpotentialI}, \ref{GpotentialII}, \ref{GpotentialIII} taken from \citet{Faucher2006ApJ}}
\begin{tabular}{c c c c c} \hline\hline
Constant & Disk-Halo (dh) & Bulge (b) & Nucleus \\\hline
M & 145.0 $\times 10^9$ $M_\odot$ & 9.3 $\times 10^9$ $M_\odot$ & 10.0 $\times 10^9$ $M_\odot$ \\
$\beta_1$ & 0.4 & & \\
$\beta_2$ & 0.5 & & \\
$\beta_3$ & 0.1 & & \\
$h_1$ & 0.325 kpc & & & \\
$h_2$ & 0.090 kpc & & & \\
$h_3$ & 0.125 kpc & & & \\
$a_G$ & 2.4 kpc & & & \\
$b$ & 5.5 kpc & 0.25 kpc & 1.5 kpc & \\\hline
\end{tabular}
\end{table}

\section{THE NEAREST SUPERNOVAE}
\label{Sec5}
\begin{figure}
\centering
\includegraphics[width=64mm]{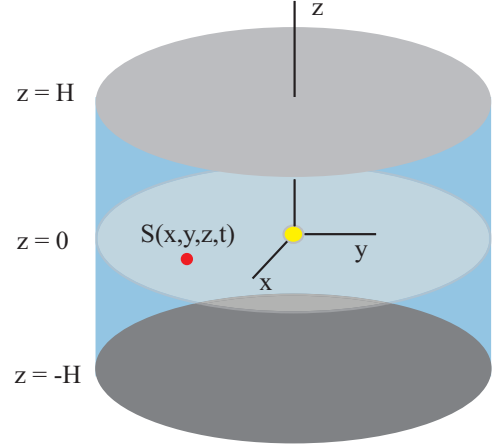}
\caption{The geometry used to simulate the GCR flux in the solar neighbourhood. The Solar System is the yellow dot in the centre and $x$ and $y$ axis are in the Galactic plane. Any GCR flux originating from supernovae in the ISM (blue) is assumed to escape from the ISM at the two planes $z=H$ and $z=-H$, where the GCR flux is therefore set to zero. The red dot indicates the position of one source of GCR.}
\label{Geometry}
\end{figure}

The averaging of supernova (SN) rates in bins of 8 Myr, as in Fig. \ref{SN_WEBDAclusters_SOLAR}c, provides a low-resolution overview of substantial changes in the background of GCR reaching the Solar System during the past 500 Myr. But at a high resolution, not directly available from the open star cluster data, one would see much larger short-lived fluctuations in GCR due to the nearest supernovae. The main purpose in the following is to demonstrate the type of fluctuations in GCR expected at Earth as a function of time. Since the most important energies for ionization in the lower atmosphere are 10-20 GeV, GCR's with energies of 10 GeV will be used. A model is needed but, with today's computers, elaborate models of GCR propagation with finite grids cannot resolve the steep gradients around point-like sources \citep{Bushing2005ApJ}. The approach used here is therefore to solve the transport equation analytically, although that means the model has to be simple. In this section a model of the temporal variation of GCR at the Solar System will be calculated, considering only the transport of GCR by diffusion in the interstellar medium (ISM), assuming it to be homogeneous for a radius of 1 kpc in the Galactic plane centred on the Solar System, and with a Galactic halo height of about 3-4 kpc beyond which particles are presumed lost. The effect of GCR modulation by solar activity and the Earth's magnetosphere are not included since they are of second order (10\%) compared to the SN induced variations for 10 GeV particles. Figure \ref{Geometry} displays the geometry. The transport of GCR in the ISM is simplified to a diffusion equation of the form, following \citet{Busching2004,Bushing2005ApJ},
\begin{figure}
\centering
\includegraphics[width=84mm]{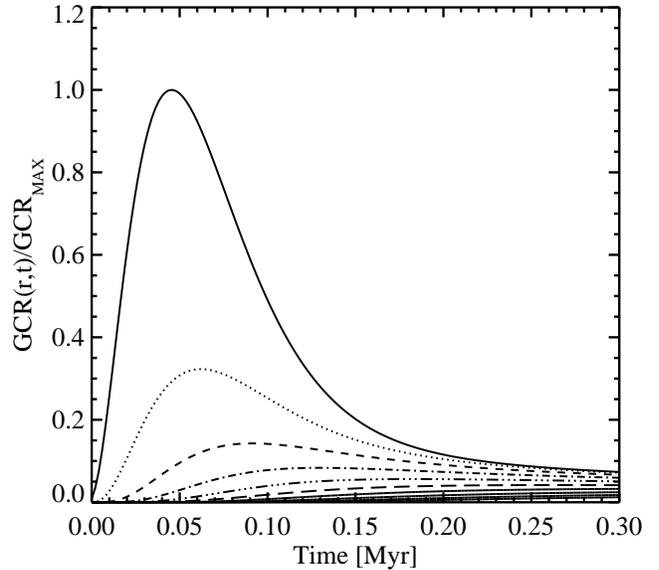}
\caption{\label{single_SN} The GCR flux (10 GeV) enhancement caused by individual SN events, located in the Galactic plane, at increasing distances from the Earth. The solid curve is for 10 pc and subsequent curves are for distances 110, 210 ... 910 pc, and a height of the ISM of $2H=3$ kpc.}
\end{figure}
\begin{figure}
\centering
\includegraphics[width=84mm]{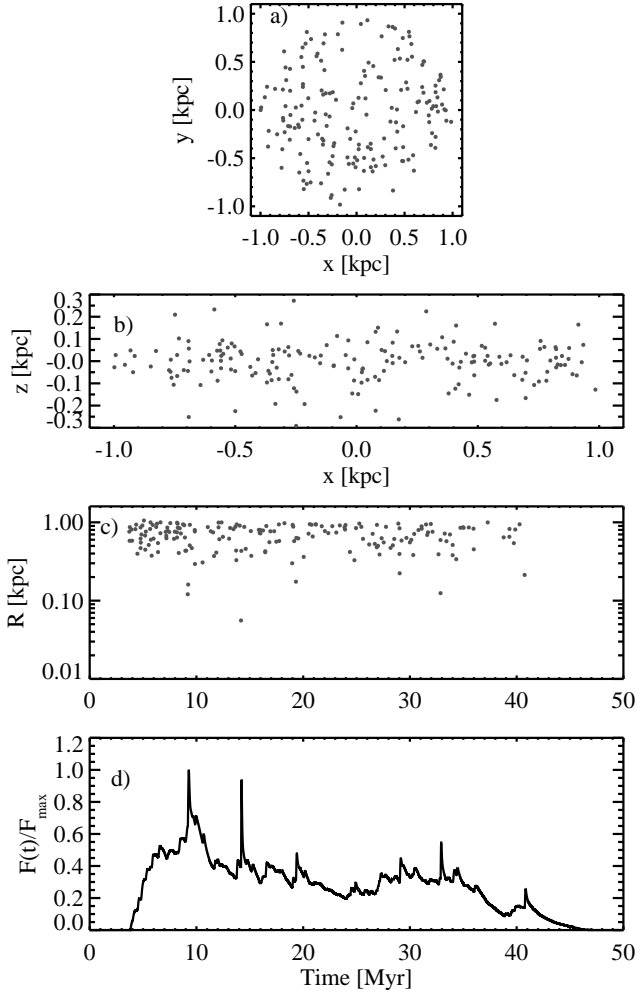}
\caption{\label{GCR_fluc_respons} Model of the frequency of supernovae resulting from a period of $\Delta t$ = 8 Myr of initial star formation. \textbf{a)} is a uniform distribution of SNs in the Galactic plane (x,y) for distances $r < 1$kpc. \textbf{b)} is the distribution of SNs perpendicular to the Galactic plane with a scale height of 90 pc \citep{MillerScalo1979ApJS}. The height of the ISM is $2H=3$ kpc. \textbf{c)} is the temporal distribution of SNs on a scale of 50 Myr, derived from the SN response function, as a function of the distance from the Solar System. Finally \textbf{d)} is the variation in GCR (10 GeV) caused by star formation in the initial $\Delta t$ = 8 Myr interval. The tallest spikes in \textbf{d)} correspond with the nearest SNs in \textbf{c)}. The GCR curve is scaled to the present SN rate = 27 kpc$^{-2}$Myr$^{-1}$ \citep{Grenier2000AA}. }
\end{figure}
\begin{equation}\label{diffusioneq}
\frac{\partial N(P,x,y,z,t)}{\partial t} = S(x_0,y_0,z_0,t) + k(P)\nabla^2 N(P,x,y,z,t)
\end{equation}
where $N(P,x,y,z,t)$ is the density of GCR particles with rigidity $P$, $S(x_0,y_0,z_0,t)$ is a source term of GCR particles from a supernova remnant, and $k(P)$ is the diffusion constant given by
\begin{equation}\label{k}
k(P) = k_0 \left(\frac{P}{P_0} \right)^{0.6}
\end{equation}
where $k_0=0.26$ kpc$^2$Myr$^{-1}$ and $P_0=$ 4GV/c \citep{Ginzburg1980,Bushing2005ApJ,Dimitrakoudis2009APh}.
Since the interest is in the GCR flux in the solar neighbourhood, the geometry will be that of a slab with height $2H$, and infinite extent in the perpendicular directions, in a cartesian $x, y, z$ coordinate system as in Fig. \ref{Geometry}. The slab is a simple model of the ISM. So the boundary conditions are
\begin{eqnarray}\label{BC}
\nonumber
N(P,x,y,z=\pm H,t) &=& 0,\\\nonumber ~~ -\infty <x < \infty, ~~ -\infty <y < &\infty&, ~~ -H \leq z \leq H \\
\end{eqnarray}
The source function of GCR from one SN occurring at $(x_i,y_i,z_i,t_i)$ \citep{Bushing2005ApJ}, is written as
\begin{eqnarray}\nonumber
S(P,x,y,z,t,x_i,y_i,z_i,t_i) =&& \\\nonumber
(t-t_i) \exp \left(-\frac{t-t_i}{\tau}\right) \left(\frac{P}{P_0} \right)^{-\alpha} \Theta(t-t_i) \times&& \\ \delta(x-x_i)\delta(y-y_i)\delta(z-z_i)&&
\end{eqnarray}
where $\Theta(t)$ is the unit step function, and $\delta$ is the delta function, and $\alpha$ is the slope of the source spectrum. The Green's function for the diffusion equation for a slab geometry is found from
\begin{eqnarray}\label{Geq}\nonumber
\frac{\partial G(x,y,z,x',y',z',t,t')}{\partial t}-k\nabla^2 G(x,y,z,x',y',z',t,t') &=& \\ \delta(x-x')\delta(y-y')\delta(z-z')\delta(t-t')&&
\end{eqnarray}
and is
\begin{eqnarray}
\label{GREENSOL} \nonumber
G(x,y,z,x',y',z',t,t') =&    & \\\nonumber
\frac{1}{4 \pi k H (t-t')} \exp \left[- \frac{(x-x')^2+(y-y')^2}{4 k (t-t')} \right] \times&&\\ \nonumber
\left( \sum_{n=1}^{\infty}\right. \exp\left[- \left(\frac{\pi n}{2H}\right)^2 k (t-t') \right]\times \\ \left.
\sin \left[\frac{\pi n(z+H)}{2H}\right] \sin \left[\frac{\pi n (z'+H)}{2H}\right] \right)&&
\end{eqnarray}
The solution to the original equation becomes
\begin{eqnarray}\label{Solfinal}\nonumber
N(x,y,z,t) &=&\\\nonumber \int_{-\infty}^{\infty}\int_{-\infty}^{\infty}\int_{-H}^{H}\int_{t'}^{t} G(x,y,z,x',y',z',t,t')\times&& \\ S(x',y',z',t') dx'dy'dz'dt'&&
\end{eqnarray}

Figure \ref{single_SN} shows an example for the above solution where the enhancement of the GCR flux (10 GeV) caused by an individual SN, located in the Galactic plane, at increasing distances from the Earth. It is then possible to simulate the effect of a large number of SNs occurring at different times by simple superposition of the solutions in Eq. \ref{Solfinal}. Assuming that the density of SNs varies in accordance with the observed changes in the SN rates as calculated from the open clusters, as in Fig. \ref{SN_WEBDAclusters_SOLAR}c, one can simulate a varying GCR flux.
The procedure is as follows:
\begin{itemize}
\item Use the calculated cluster production rate during the last 500 Myr (as shown in Fig. \ref{SN_WEBDAclusters_SOLAR}b), as the basis of the stochastic calculation.
\item Estimate the number of massive stars that go supernova in a time step of $\Delta t = 8$ Myr by scaling SN(0) to the present rate of SN in the 1 kpc region, as SN(0) = 27 kpc$^{-2}$Myr$^{-1}$ $\pi$ 1 kpc$^2$ 8 Myr = 679 SN \citep{Grenier2000AA}.
\item Distribute the SN in space by using a random generator to make a uniform distribution (x,y) in the Galactic plane in the 1 kpc region, and an exponential distribution with a scale height of 90 pc \citep{MillerScalo1979ApJS} around the alactic plane with coordinate $z$. Figs. \ref{GCR_fluc_respons}a and \ref{GCR_fluc_respons}b show one realization of such a spatial distribution based on an 8 Myr time step.
\item Having the estimated number of massive stars (the SN progenitors) in a time step $\Delta t$, use the response function shown in Fig. \ref{SNresponsfunction_II} to calculate a temporal probability distribution prescribing the times $t_i$ when the massive stars go SN. Figure \ref{GCR_fluc_respons}c shows the temporal distribution and the distances of the SNs from the Solar System in the following $\approx$ 40 Myr.
\item From the temporal and spatial distribution of SNs, calculate the solution in Eq. \ref{Solfinal} for each SN at distance and time $(r_{i},t_i)$, and add the solutions to obtain the temporal GCR flux at the Solar System.
\end{itemize}
Figure \ref{GCR_fluc_respons}d shows the temporal evolution based on the above procedure from a single time step $\Delta t$ = 8 Myr. The massive SN progenitors stars go SN with the coordinates $(x_i,y_i,z_i,t_i)$ within a 1 kpc distance in the solar neighbourhood and within an approximately 40 Myr period following the initial formation of the massive stars. For each SN with a distance and a time the solution based on Eqn. \ref{Solfinal} is calculated, and then all the solutions are added to give the temporal variation at the position of the Solar System. As seen in Fig. \ref{GCR_fluc_respons}c and d only a relatively close SN results in a clear spike in the GCR flux.
\begin{figure}
\centering
\includegraphics[width=84mm]{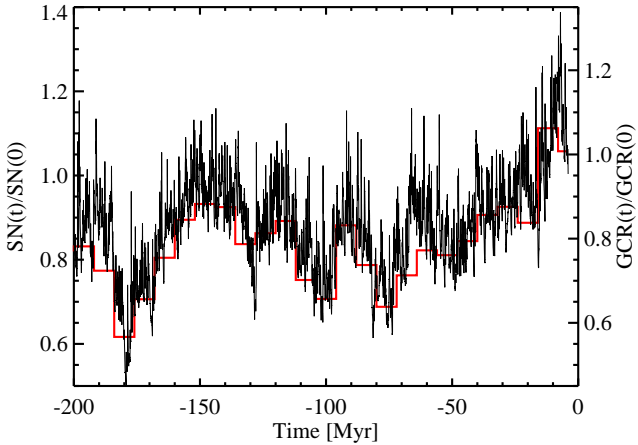}
\caption{\label{GCR(t)} Variability in the GCR flux (10 GeV) to the Solar System (red curve) calculated over 200 Myr using a stochastic model of SN distribution (see text). The red curve shows the SN rate in bins of 8 Myr intervals. The height of the ISM is $2H=3$ kpc.}
\end{figure}
Figure \ref{GCR(t)} illustrates one (random) realization of a calculation of the GCR flux caused by the varying SN rate during the past 200 Myr as determined from the estimate of open clusters production (Fig. \ref{SN_WEBDAclusters_SOLAR}b). Notice that the GCR variation is highly fluctuating with many positive excursions caused by SNs near the Solar System. The variations in GCR shown here are calculated for 10 GeV particles and $k_0 =$ 0.026 kpc$^2$Myr$^{-1}$. With respect to changes in ionization of the terrestrial atmosphere, these are expected to be much larger than for similar variations in the 10 GeV flux caused by solar activity. The reason is that solar activity can modulate GCR primary particles only at the low end of the energy range, and leaves those $>$80 GeV almost unaffected. A close SN delivers the whole spectrum up to 10$^{17}$ GeV. This effect of high-energy primaries on atmospheric ionization will be described in more detail in a forthcoming paper (Svensmark, in preparation).

Very close SNs, at $r <$ 10 pc, seem to be extremely rare, in agreement with previous estimates \citep{Fields1999NewA}.
Unfortunately no terrestrial archives can directly provide a time series to test this high-frequency variability of the GCR flux.
But if the impact of GCR variations is significant for the Earth's climate, consequences of the brief spikes can be sought in the geological record, as in the following section.

\section{RESULTS: LOCAL SUPERNOVA HISTORY AND CHANGES OF CLIMATE}
\label{Sec6}
As noted in the Sect. \ref{intro}, evidence has accumulated in recent years that the influx of Galactic cosmic rays, as modulated by solar magnetic activity, influences the Earth's climate. Climate variations during the past 10,000 years, detected in deep-sea cores or in stalagmites from caves, closely tracked changes in the GCR flux as recorded by the radiogenic isotopes $^{10}$Be and $^{14}$C \citep{Bond2001,Neff2001}. A proposed mechanism by which GCR can affect the Earth's climate is via an influence on cloudiness \citep{ne59,di75,Svensmark1997JATP,Svensmark1998PhRvL,Svensmark2000PhRvL,ha05}. The basic assumption is that negative ions created by GCR help to make fine aerosols which subsequently grow to cloud condensation nuclei on which water droplets form. As a result more GCR create more low clouds and the larger albedo results in a cooling of the Earth. Although thin clouds at higher altitudes can exert a warming on Earths surface the net effect of all clouds is a cooling \citep{Hartmann1993}. In addition observations suggest that high clouds are unaffected by variations in GCR \citep{Svensmark2000PhRvL}.

In experimental chambers that simulate atmospheric conditions, negative ions produce new ultra-fine aerosols 2-3 nm in diameter \citep{Svensmark2007RSPSA,Kirkby2011}. To boost cloud formation, these additional ion-generated aerosols have to grow to sizes larger than 50 nm to become cloud condensation nuclei (CCN).  An increased competition for condensable vapours, especially sulphuric acid produced by UV-photochemical reactions involving ozone, sulphur dioxide and water, might slow down the growth so much that any small increase in aerosols would be lost before reaching CCN size \citep{Snow-Kropla2011}. A recent experiment seems to have resolved this conundrum, by finding a second ion-induced pathway for the formation of sulphuric acid \citep{Enghoff2012,Svensmark2012}. As sulphuric acid is one of the most important molecules in the formation and growth of atmospheric aerosols, its production by GCR ionization appears to give a very simple explanation of how GCR changes can control the number of CCN, which ultimately affects the climate on Earth.

Moreover, the whole chain of effects from solar activity to cosmic ray ionization to aerosols and liquid-water clouds is discernible in the real atmosphere during the days following explosive coronal mass ejections that reduce the GCR flux near Earth \citep{Svensmark2009GeoRL,JSvensmark2012}. For the most influential recent events, the liquid water in the oceanic atmosphere decreased by as much as 7\%.
Over decades, centuries and millennia, the GCR flux reaching the Solar System from the Galactic environment is more or less constant. Changes in the flux at the Earth have been due to changing solar activity, with variations of the order 10\%. But on longer time scales (Myr) the changes in local star formation rate and subsequent SN and GCR acceleration can produce much larger changes in the GCR flux. Effects of variations of GCR due to the stars rather than the Sun offer an independent test of the proposed link between GCR and Earth's climate \citep{Shaviv2002PRL,Shaviv2003NewA,Shaviv2004GSA,Marcos2004NewA}.

\begin{figure}
\centering
\includegraphics[width=84mm]{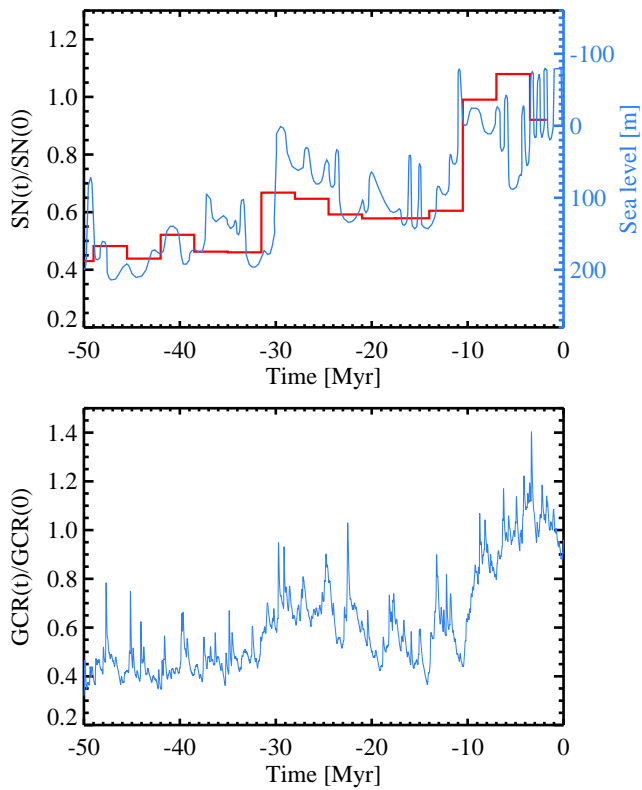}
\caption{\label{Sealevel_SN} (a) Short-lived falls in sea level (marine regressions) seen in the stratigraphic record (data of \citet{Haq1987} obtained from \citet{Miller2005}) during the past 50 Myr are suggested to be signals of the nearest supernovae, enhancing GCR and provoking increased cloudiness and glaciation. The blue curve in the top panel, where the right-hand scale is inverted, shows an overall fall in global sea level due mainly to the loss of ocean water into ice sheets of increasing volume grounded in the polar regions. The trend corresponds broadly to the overall SN rate (red curve), here presented in bins of 3.5 Myr. Because of the scale inversion, the short-lived marine regressions appear as brief spikes superimposed on the general trend in sea level. In variance and frequency they resemble the excursions in GCR modelled in Fig. \ref{GCR(t)}, due to the nearest supernovae. In the lower panel the simulated GCR flux (10 GeV) over the past 50 Myr incorporates excursions due to the nearest individual supernovae. Note that this is just one random realization of the statistics of SN events.}
\end{figure}

When the GCR flux increases as a result of astrophysical activity, the expected increases in cloudiness and the Earth's albedo should cause the climate to cool. Changes of climate during the history of the Earth, as detected by geologists, show big swings between warm and cold conditions, and many occurrences of continental ice sheets. In this perspective, the sudden large increases in the flux due to the nearest supernovae, considered in Sect. \ref{Sec5}, should result in severe global cooling events on time scales of the order of 10,000-100,000 years, which are long compared with the life cycles of species but quick in geological terms. When one searches the geological record for symptoms of brief but severe cooling events with the magnitude, time scale and frequency appropriate for signals of the nearest SNs, the most promising are short-lived falls in global sea level, called marine regressions, for which there exists no other satisfactorily comprehensive explanation.

By exposing beaches to erosion, the marine regressions have left signatures of discontinuous strata that are used routinely for seismic stratigraphy. In the decades since they came centre-stage in geophysics \citep{Haq1987}, hypotheses on offer to explain the short-term falls in sea level have included trapping of water in lakes, rifts or underground (hydro-eustasy) and changes in the supply of sediments to beaches possibly linked to 400-kyr insolation cycles ("supra-Milankovitch" cycles). Some of the smaller regressions may be explicable in such ways, but many falls $>$25 m seem to require the presence of ephemeral ice sheets, even during the warmest geological periods \citep{Miller2005}. Excellent support for the regressions history and a link to glaciations comes from \citet{Billups2002} who used Mg/Ca and oxygen isotopes in benthic foraminifera to assess changes over the past 27 Myr.

In the absence of other evidence for when the nearest SN events occurred, the hypothesis that they provoked major marine regressions can be considered by referring to the fluctuations in GCR calculated in Sec \ref{Sec5}. Fig. \ref{Sealevel_SN} focuses on the sea level changes over the past 50 Myr, shown in the top panel (scale inverted). Also in the top panel is the varying SN rate (red curve), shown in this case with a resolution of 3.5 Myr. The overall covariation of sea levels and SN rates is good, but it becomes better in detail when an arbitrary randomization of the SN events within each period of 3.5 Myr generates the GCR flux shown in the lower panel. The count of major marine regressions includes $\approx$ 25 regressions of $>$25 m, which can be compared with an expectation of $\approx$ 25 major GCR excursions due to the nearest supernova, as shown in the lower panel of Fig. \ref{Sealevel_SN} ($\approx$ 0.5 Myr$^{-1}$). The aim here is not to try to achieve a perfect covariance by further statistical iterations, but to illustrate that, in the absence of any other explanation for them, the fast sea level falls are just what are to be expected as signals of closer-than-usual SN detonations.

The more persistent changes in sea level seen in Fig. \ref{Sealevel_SN} tell a grander story about the build-up of ice sheets on Antarctica, as the Earth began to develop its present glacial climate. It was a hesitant process \citep{Zachos2001} with major cooling and glaciation beginning 33 Myr ago, a warmer interval starting 26 Myr ago, and refreezing in the gradual build-up to recent ice volumes beginning 10-14 Myr ago. It can be seen from the red curve in the upper panel that this timetable is broadly consistent with ups and downs in the SN rate.

\begin{figure}
\centering
\includegraphics[width=84mm]{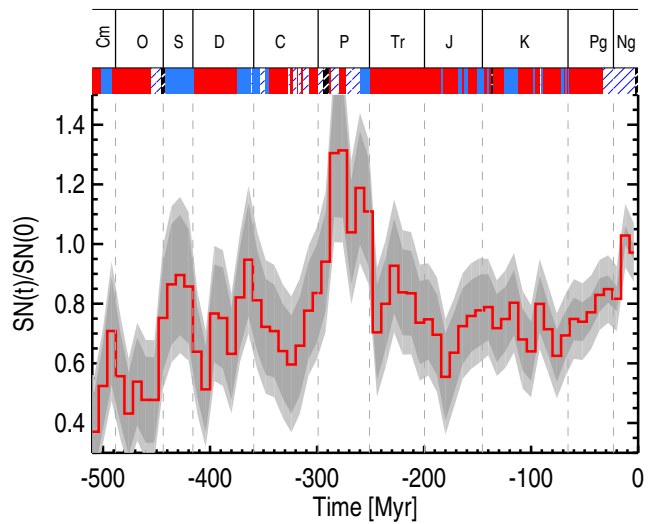}
\caption{\label{WEBDAcluster_CLIMATE} Variations in SN rates during the past 500 Myr (red curve) together with a $\pm$1-$\sigma$ uncertainty (dark grey band) and $\pm$1-$\sigma$ uncertainty including Poisson noise (light grey band). The vertical dashed lines are the separation between geological periods. The coloured band indicates climatic periods as given in Table 3: warm periods (red), cold periods (blue), glacial periods (white and blue hatched bars), and finally peak glaciations (black and white hatched bars). Notice the correspondence between high SN activity and cold/glacial climate. Abbreviations for geological periods are: Cm, Cambrian; O, Ordovician; S, Silurian; D, Devonian; C, Carboniferous; P, Permian; Tr, Triassic; J, Jurassic; K, Cretaceous; Pg, Palaeogene; Ng, Neogene. Plot starts at -510 Myr.}
\end{figure}

Reverting to the overall pattern of SN rates derived in Sect. \ref{Sec3}, these can be compared with more gradual but larger changes in climate in the past 500 Myr, alternating between persistently warm and persistently cold conditions. Cold and glacial intervals are listed in Table 3 and shown in the coloured band in Fig. \ref{WEBDAcluster_CLIMATE}. All geological periods and dates are given with reference to the Geological Time Scale 2004 of \citet{Gradstein2005}. Starting with data from \citet{Royer2006} and \citet{Saltzman2005}, a scan of the literature adds important details. Also noted are what seem to be the "peak ice" times over long intervals, which are of interest for the GCR hypothesis, although one should be aware that the magnitudes of those glaciations differed greatly.

Despite the uncertainties about peering so far back in time, both astrophysically and geologically, the association between cold conditions and high SN rates stands out clearly in Fig. \ref{WEBDAcluster_CLIMATE} and provides strong support for a very long-term GCR-climate link that is independent of solar variations. Also noticeable in Fig. \ref{WEBDAcluster_CLIMATE} is the way most geological periods fit neatly around either an upturn or a downturn the SN rates, suggesting that the transition from one period to the next is connected to a major change in the astrophysical environment.

\begin{figure}
\centering
\includegraphics[width=84mm]{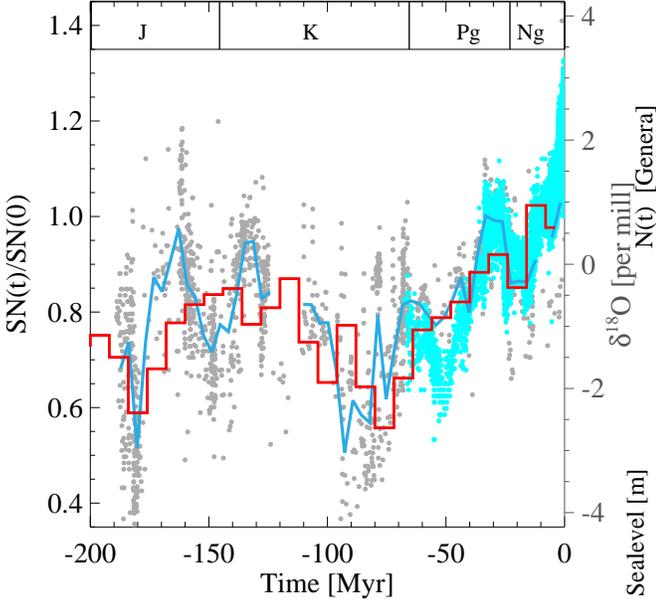}
\caption{\label{SN_d18O_200Ma} Measurements of $\delta^{18}$O over the last 200 Myr (coloured points) compared with SN history over the same period (red curve). Light blue data points are from \citet{Zachos2001} and grey circles are from \citet{Veizer2008}. The dark blue curve is an average of the grey points. The light blue points have been offset by -2.5 per mil relative to the grey data to take account of the provenance of the data from deeper water.}
\end{figure}

Widely used as a quantitative measure of changes in temperature is the ratio between the two stable isotopes of oxygen, $^{18}$O and $^{16}$O, found in sediments and fossil sea shells. Oceanic enrichment with $^{18}$O occurs in cold conditions for physical reasons, of which the simplest is that light water molecules with pure $^{16}$O evaporate a little more readily. $\delta^{18}$O is measured in parts per mill (i.e. parts per thousand) compared with a reference material (Vienna PDB, -2.2 per mill) and it is important to be aware of the provenance of the data, because regional differences can obscure the global picture. That is why, in comparing $\delta^{18}$O with SN rates over the past 200 million years, in Fig. \ref{SN_d18O_200Ma}, prominence is given to results from carefully chosen mid-latitude sea shells \citep{Zachos2001}, although other sources are also included \citep{Veizer2008}. As one can see in Fig. \ref{SN_d18O_200Ma} there seems to be, with some exceptions, an overall correspondence between $\delta^{18}$O and SN rates.

\begin{table*}
\caption{Episodes of glaciation and cold intervals as displayed in Fig. \ref{WEBDAcluster_CLIMATE}. {(1) \citet{Saltzman2005}; (2) \citet{Saltzman2005II}; (3) \citet{Trotter2008}; (4) \citet{Royer2006}; (5) \citet{FIELDING2008}; (6) \citet{McArthur2007}; (7) \citet{Bornemann2008Sci}; (8) \citet{Baumann1999}}}
\begin{tabular}{|c|c|c|c|l|l|} \hline\hline
& Period & Myr BP & Cold intervals&Conditions & Reference \\\hline
Cm & Cambrian & 542 & 502 -492 & Cold & 1 \\
O & Ordovician &488.3& 455.0 -443.7 & Glacial& 2 \\
& & & 450 -445 & Cold& 1 \\
& & & 445 -442 &Peak ice& 3 \\
S & Silurian &443.7& 442 -415 & Cold& 1 \\
D & Devonian &416 & 375 -361.4 & Cold& 1 \\
& & & 361.4 -360.6 & Glacial& 4 \\
& & & 360.6 -359.2 & Cold& 1 \\
C & Carboniferous& 359.2 & 375 -353 & Cold& 1 \\
& & & 353 -349 & Glacial& 4 \\
& & & 349 -345 & Cold& 1 \\
& & & 326.5 -325.5 & Glacial& 5 \\
& & & 322.5 -319.5 & Glacial& 5 \\
& & & 317.0 -315.0 & Glacial& 5 \\
P & Permian & 299& 299 -294 & Glacial& 5 \\
& & & 294 -291 &Peak ice& 5 \\
& & & 287 -280 & Glacial& 5 \\
& & & 273 -268 & Glacial& 5 \\
& & & 267 -260 & Glacial& 5 \\
& & & 260 -251 & Cold& 4 \\
Tr & Triassic &251 & & & \\
J & Jurassic &199.6& 184 -183 & Cold& 4 \\
& & & 167.7 -164.7& Cold& 4 \\
& & & 162 -159 & Cold& 4 \\
& & & 150 -144 & Cold& 4 \\
K & Cretaceous &145.5& 140.5 -139.5 & Cold& 4 \\
& & & 137.5 -137 & Cold& 4 \\
& & & 137 -136 &Peak ice& 6 \\
& & & 125.0 -112.0& Cold& 4 \\
& & & 97.5 -96.5 & Cold& 4 \\
& & & 91.3 -91.1 & Glacial& 7 \\
& & & 91.1 -89 & Cold& 4 \\
& & & 71.6 -69.6 & Cold& 4 \\
& & & 67.5 -66.5 & Cold& 4 \\
Pg & Palaeogene &65.5& 65.6 -65 & Cold& 4 \\
& & & 33 -23.03 & Glacial& 4 \\
Ng & Neogene & 23.03& 23.03 -2.8 & Glacial& 4 \\
& & & 2.8 -0 &Peak ice& 8 \\\hline
\end{tabular}
\end{table*}

\section{RESULTS: ECOLOGICAL CORRELATIONS WITH SUPERNOVA HISTORY}
\label{Sec7}

\begin{figure}
\centering
\includegraphics[width=84mm]{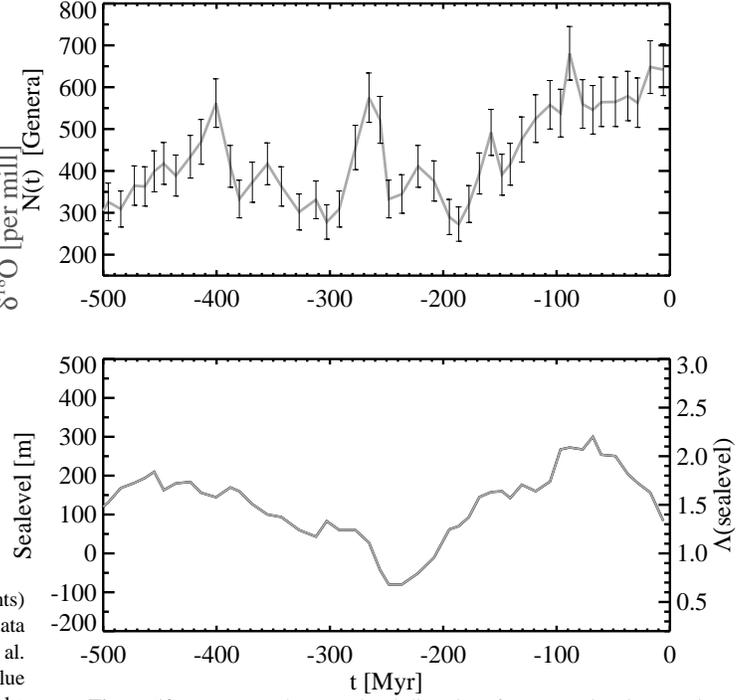}
\caption{\label{GeneraSealevel} Upper panel: genus-level diversity of extant and extinct marine invertebrates as an index of variable biodiversity (\citet{Alroy2008}, based on a sampling-standardized analysis of the Paleobiology Database). Lower panel: First-order sea level variations attributable to tectonics (adapted from \citet{Haq1987,Haq2008}).}
\end{figure}
\begin{figure}
\centering
\includegraphics[width=84mm]{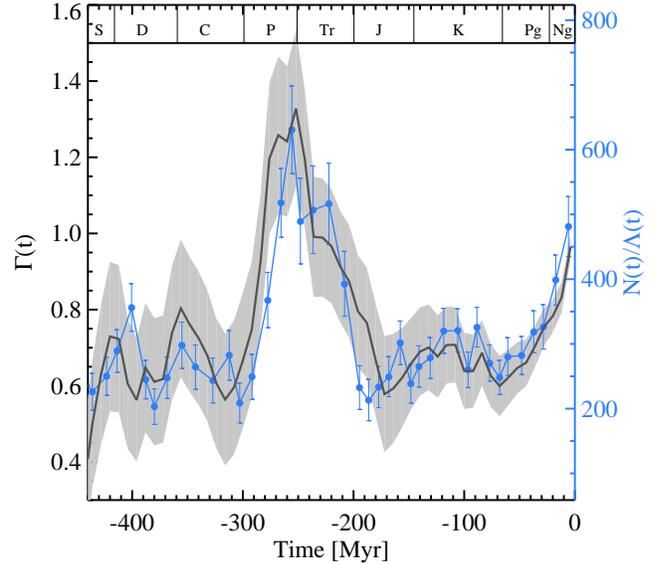}
\caption{\label{GeneraAlroy} SN history and marine invertebrate genera. The black curve is based on the SN rates and given by Eq. \ref{Genera3}. The blue curve shows marine genera normalized with the sea level (Eq. \ref{sealevel}) and defined as the right-hand side of Eq. \ref{Genera4}. The grey area is the 1-$\sigma$ variance calculated from a Monte Carlo simulation. The error bars on the genera (blue curve) show a minimum 1-$\sigma$ uncertainty since an error estimate is not available for the sea level data. Evidently marine biodiversity is largely explained by a combination of sea level and astrophysical activity.}
\end{figure}
One index of success or setback for life in general is biodiversity. The evolution of living creatures can be gauged by the rate of origination of new forms (classified, for example, as orders, families, genera or species) and the rate of extinction of old forms. The difference at any stage results in an increase or decrease of global biodiversity. By counting known fossils, palaeontologists have found large variations in marine biodiversity over the past 500 Myr.
The top panel of Fig. \ref{GeneraSealevel} shows the recently re-assessed genus-level diversity of marine invertebrate animals. One of the many intrinsic and environmental factors suspected of influencing biodiversity is sea level, shown in the lower panel of Fig. \ref{GeneraSealevel}. Excluding the brief marine regressions discussed earlier, this curve is the first-order envelope of long-term sea level. Conventionally, in plate tectonics, the accretion of mid-ocean ridges during sea-floor spreading pushes up the sea level \citep{Russell1968}. The long-term sea level minimum 200 Myr ago coincided with the completion of the supercontinent of Pangaea, whilst the peaks in sea level 100-70 Myr and 450 Myr ago were associated with the spreading of new oceans following the breakup of Pangaea and of its predecessor Pannotia.

A higher sea level will result in flooding of low inland areas, and increase the total length of coastline and the area of the continental shelves. This results in more heterogeneous habitats in which species can evolve, leading to an increase in diversity, and \citet{Miller2005} offer this as the reason for the development of the three eukaryotic phytoplankton clades (lineages) that dominate the modern ocean. But the rough correspondence between marine invertebrate diversity and sea level seen in Fig. \ref{GeneraSealevel} is plainly not the whole story.
Here the novel assumption is that variations in SN rates will also change conditions for the evolution of life by changing the climate and thereby altering, for example, the circulation of nutrients and the variety of habitats between the Equator and the poles. If the sea level and the changes in climate caused by SN are both important one would expect that the temporal evolution of marine invertebrate diversity $\textsf{N}(t)$ should be a function of SN rate $\Gamma(\textrm{SN},t)$ and sea level $\Lambda(Sea\b{ }level,t)$ written as
\begin{equation}
\label{Genera1}
\textsf{N}(t) = \Gamma(\textrm{SN},t) \Lambda({{Sea\b{ }level}},t) + \epsilon(t)
\end{equation}
where $\epsilon(t)$ is a hopefully small noise term. If the genera count is normalized by the sea level, a possible effect of the astrophysics will be as
\begin{equation}
\label{Genera2}
\Gamma(\textrm{SN},t) = \frac{\textsf{N}(t)}{\Lambda({{Sea\b{ }level}},t)} + \epsilon(t)
\end{equation}
where $\epsilon(t)$ again is a noise term. Finally, since a mass extinction can require 10 - 40 Myr recovery time \citep{Alroy2008PNAS}, it is further to be expected that a large change in SN can also lead to a persistent effect on the genera count. The function $\Gamma(\textrm{SN},t)$ should therefore also depend on earlier times and the simplest relation is
\begin{equation}
\label{Genera3}
\Gamma(\textrm{SN},t) = \nu_1 \int_{-\infty}^t SN(t')\exp{\left[-\lambda (t-t')\right]} dt' + \nu_2
\end{equation}
where $\nu_1$ and $\nu_2$ are constants.
Inserting Eq. \ref{Genera3} in Eq. \ref{Genera2} one gets
\begin{eqnarray}
\label{Genera4}\nonumber
\frac{\textsf{N}(t)}{\Lambda({{Sea\b{ }level}},t)} &=&\\ \nu_1 \int_{-\infty}^t SN(t')\exp{\left[-\lambda (t-t')\right]} dt' &+& \nu_2 + \epsilon(t)
\end{eqnarray}
Notice that the left hand side depends only on terrestrial quantities and the right hand side depends only on astrophysical quantities.
Finally the sea level function is approximated by the two first terms of a Taylor expansion of $\Lambda({{Sea\b{ }level}},t)$ as
\begin{equation}
\label{sealevel}
\Lambda({{Sea\b{ }level}},t) = \alpha + \beta ~ {Sea\b{ }level(t)}
\end{equation}
where $\alpha$ and $\beta$ are constants. Figure \ref{GeneraAlroy} shows, as the blue curve, the left-hand side of Eq. \ref{Genera4} which is the ratio between $\textsf{N}(t)$ and $\Lambda({{Sea\b{ }level}},t)$. The black curve in Fig. \ref{GeneraAlroy} is the right-hand side of Eq. (\ref{Genera4}). The curves are displayed for $\alpha = 1$, $\beta = 1/250$ m$^{-1}$ and $\lambda =$ 1/20 Myr$^{-1}$. These values ensure the best agreement between the astrophysical part (right-hand side of Eq. (\ref{Genera4})) and the terrestrial part (the left-hand side of Eq. (\ref{Genera4})). The correlation coefficient is 0.89 for the last 400 Myr. Notice the overall agreement between the blue and black curve, which gives confidence in the basic assumption made in Eq. \ref{Genera1}, namely that the chief governors of marine invertebrate diversity are tectonics (sea level) and astrophysics (local SN rate).

Another global index of life's successes and setbacks, to consider alongside biodiversity, is its primary productivity. This is the rate at which all the world's carbon-fixing micro-organisms and plants, powered typically by sunlight, assimilate carbon for their own growth and for the eventual nourishment of animals and fungi. As the motor of the carbon cycle, the production of organic material removes carbon dioxide from oceanic water and the atmosphere, and replaces it with oxygen.
A simple working hypothesis, suggested by carbon-isotope data for the past 4 Gyr \citep{Svensmark2006ANII}, is that primary productivity increases in glacial conditions, perhaps because of better nutrient supplies, caused by a more vigorous mixing in the oceans during cold conditions. This hypothesis (to be discussed later) would predict:
\begin{enumerate}
\item A drawdown of CO$_{2}$ from the environment in glacial conditions.
Since organic productivity consumes CO$_2$ there should be an impact on the levels of atmospheric and oceanic CO$_2$. High productivity draws down CO$_2$, until ultimately the productivity rise is halted not only by exhaustion of nutrients but by the scarcity of CO$_2$, which should prevent a total loss of environmental CO$_2$. Conversely, low productivity should result in an accumulation of underemployed CO$_2$.
\item Due to the increased organic productivity an increase in the heavy stable isotope of carbon, $^{13}$C, is expected in the oceans during glacial conditions.
\end{enumerate}
\begin{figure}
\centering
\includegraphics[width=84mm]{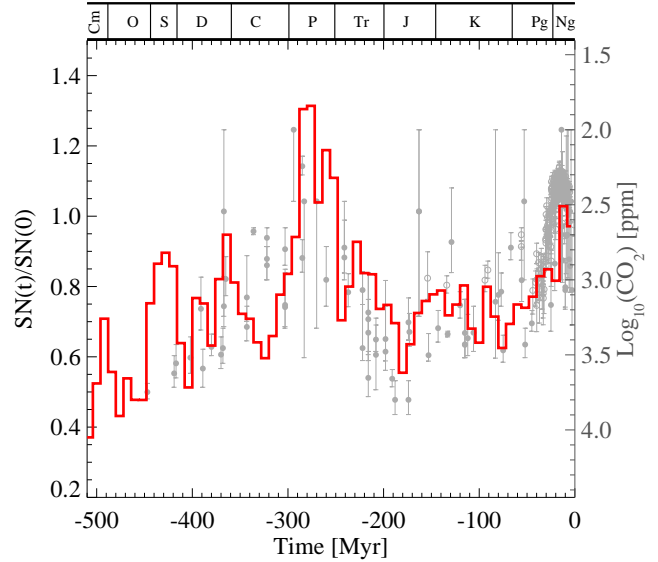}
\caption{\label{GEOCO2} Used here as a proxy for Galactic cosmic rays (GCR) reaching the Earth, the local supernova rate (red curve) from Fig. \ref{SN_WEBDAclusters_SOLAR}c is compared with the logarithm of concentrations of CO$_2$ in the air (grey circles, scale inverted). The filled circles show CO$_2$ data from $\delta^{13}$C in palaeosols and the open circles are from foraminifera \citep{Royer2006}. The plot gives an overall impression of CO$_2$ diminishing in cold conditions associated with high SN rates, possibly because of drawdown of CO$_2$ by photosynthesis in better fertilized oceans (see text). Logarithms are used because drawdown should be self-limiting when little CO$_2$ is available. Geological periods are shown by their abbreviations as in Table 3.}
\end{figure}
\begin{figure}
\centering
\includegraphics[width=84mm]{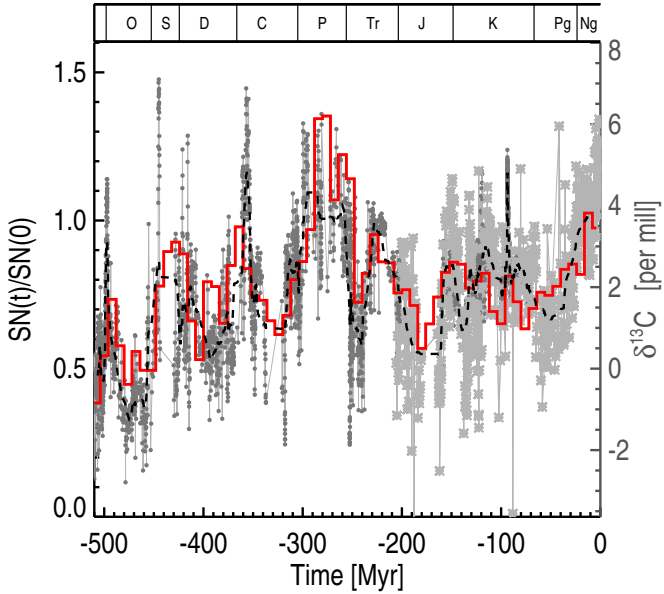}
\caption{\label{SN_delta_C13}
Changes in the global carbon cycle over the past 500 Myr are reflected in the proportions of carbon-13 in sediments ($\delta^{13}$C in parts per mill) shown by the scattered points, and compared here with the variations in the local supernova (SN) rate (red curve). Circles are $\delta^{13}$C in marine carbonates: Cambrian (not labelled) to Carboniferous C \citep{Saltzman2005}, Permian P \citep{Grossman2008}, Permo-Triassic transition P-Tr \citep{Kakuwa2006}, Triassic T \citep{Korte2005}, Jurassic J to Cretaceous K \citep{Emeis2009} and Cretaceous K \citep{JARVIS2006}. Star symbols are $\delta^{13}$C in fossil organic matter (offset by 27 per mill): Jurassic J to Neogene Ng \citep{Falkowski2005}. The black dashed line is a smoothing of the $\delta^{13}$C data. Notice that there are three brief gaps in the $\delta^{13}$C data (End-Silurian, Mid-Carboniferous and Mid-Jurassic). The plot starts at -510 Myr.
}
\end{figure}
As glacial versus warm conditions seem to follow SN rates (Figs. \ref{WEBDAcluster_CLIMATE} and \ref{SN_d18O_200Ma}) one can look for matches between CO$_2$, $^{13}$C and SN rates. In the case of CO$_2$, data are sparse in the earlier part of the 500-Myr record considered here, but more abundant later. In Fig. \ref{GEOCO2} the proxies for CO$_2$ are paleosols (fossil soils) which offer the longest time-span although less accurate at low CO$_2$ \citep{Royer2001ESR} and fossil planktonic foraminifera organisms from the oceans \citep{Royer2006}. The CO$_2$ scale is inverted because because high SN rates (cold climate) and low CO$_2$ go together, and the logarithm of the CO$_2$ concentrations is used, on the assumption that the relation is not linear, in particular when the when CO$_2$ is scarce. The match between CO$_2$ and SN rates encourages further pursuit of the hypothesis in respect of $\delta^{13}$C, as a possible indicator of primary productivity.

When organisms take in carbon to grow, mainly as CO$_2$ in photosynthesis, they prefer molecules without the rarer (1\%) $^{13}$C, and so become enriched in $^{12}$C . When they die and decay they usually return their enriched carbon to the environment, but not always. In the oceans, some of the dead phytoplankton, and other organisms that have fed on their $^{12}$C-enriched constituents, become incorporated into seabed deposits or trapped in deep water by a failure of the oceanic circulation. On land, some plants become buried as peat or coal. Organic carbon sequestration and the loss of enriched $^{12}$C leaves the global environment enriched in $^{13}$C. This becomes apparent in samples of surviving sea-floor deposits of inorganic carbonate and in the fossils of organisms that lived in the $^{13}$C-enriched environment. As the global mixing time for seawater is of the order 1000 years, the regional influences on $\delta^{13}$C should tend to average out on the million-year scale considered here.
The definition of $\delta^{13}\textrm{C}$ is given by
\begin{equation}\label{dC13}
\delta^{13}\textrm{C} (\textrm{per mill}) =\left[\frac{ \left(\frac{^{13}C}{^{12}C}\right)_{\textrm{Sample}}-1 } {\left(\frac{^{13}C}{^{12}C}\right)_{\textrm{Reference}}}\right] \times 1000
\end{equation}
where measurements are referred to $(^{13}C/^{12}C)_\textrm{reference}$ = 0.0112372 (Vienna PDB).
The carbon cycle is an interaction between different reservoirs, and its mass balance for the ocean (following \citet{Kump1999}) can be written as
\begin{equation}\label{C_reservoirs1}
\frac{dM_0}{dt} = F_\textrm{w'} - (F_\textrm{b,org} + F_\textrm{b,carb})
\end{equation}
where, $M_0$ is the change in carbon mass in the ocean atmosphere system, $F_\textrm{w'}$ is the input of carbon from weathering and volcanoes, $F_\textrm{b,org}$ is the input from organic matter and finally $F_\textrm{b,carb}$ is the carbon input from carbonate minerals.
From this the change in the oceanic (marine carbonate) carbon isotope reservoir becomes
\begin{equation}\label{C_reservoirs2}
\frac{dM_0 \delta_\textrm{carb}}{dt} = F_\textrm{w'} \delta_\textrm{w'} - F_\textrm{b,carb}\delta_\textrm{carb} - F_\textrm{b,org}( \delta_\textrm{carb} +\Delta_\textrm{B})
\end{equation}
where $\Delta_\textrm{B}= \delta_\textrm{org} -\delta_\textrm{carb} $ is the isotopic difference between organic matter and carbonate deposited in the ocean (typically of the order -25 per mil1).
On the time scales of Myr the above equations reduce to steady state ($dM_0/dt=0$), and simple algebra gives
\begin{equation}\label{C_reservoirs3}
f_\textrm{org} = \frac{F_\textrm{b,org}}{F_\textrm{w'}+F_\textrm{b,org}}= \frac{\delta_\textrm{w'}-\delta_\textrm{carb}}{\Delta_\textrm{B}}
\end{equation}
which indicates, under the assumption that the terms $\delta_\textrm{w'}$ and $\Delta_\textrm{B}$ are constant, that the term $\delta_\textrm{carb}$ is related to the fraction of carbon buried as organic matter.
Fig. \ref{SN_delta_C13} shows the variations in $^{13}$C recorded as $\delta_\textrm{carb}$ (Cambrian to Cretaceous) and $\delta_\textrm{org}$ (Jurassic to Neogene) from sources cited in the caption. The red curve is the variation in SN rate during the last 510 Myr. Despite the large variability in $\delta^{13}$C in each time interval, a dominant influence of the SN rate plainly over-rides all the complex practical and theoretical reasons why such coherence with $\delta^{13}$C might not be expected (see Sect. \ref{Sec8}).
Considered together with the climatic evidence in Figs. \ref{WEBDAcluster_CLIMATE} and \ref{SN_d18O_200Ma}, the palaeobiological Figs. \ref{GeneraAlroy}, \ref{GEOCO2}, and \ref{SN_delta_C13} give a strong impression of impacts of SN rates on both the evolution and productivity of life over the past 500 Myr. The interpretation of these empirical results is discussed below.
\section{Discussion}
\label{Sec8}

\subsection{Galactic structure and supernova rates in the Earth's vicinity (Sections 2-5)}
Given the uncertainties about the Milky Way, it seemed entirely possible at the outset of the quest for local supernova rates over 500 million years that complexities of Galactic structure, dispersal of open clusters by orbital diffusion, or insufficient data would frustrate the effort. In the outcome, the results are sufficiently coherent to give a convincing match to known spiral arms (Fig. \ref{Spiral}). At the same time they are sufficiently variegated to shed new light on Galactic structure, and in particular on big differences in structure outside and inside the solar circle seen in Figs. \ref{WEBDAclustersFIT} and \ref{WEBDAclustersINDSIDE}.

For the primary reckoning of SN rates in the solar neighbourhood, seen in Fig. \ref{SN_WEBDAclusters_SOLAR}c, the size of the region around the Solar System was chosen to be large enough for a sufficient number of clusters in the statistical ensemble, and small enough for the ensemble to be as complete as possible. As seen in Fig. \ref{Webdadist} these requirements limit the preferred radius to about 0.85 kpc. Although using a different radius does change the shape of the resulting SN curves, in the vicinity of 0.85 kpc the change is not large.
The decay law for open clusters, given in Eq. \ref{cluster2}, is a simple law that fits the data. One could argue that the decay of very young clusters is different from older clusters, warranting a decay law with two exponents that could result in a larger estimate of the number of old clusters, but the simpler approach with only one exponent in Eq. \ref{cluster2} is preferred. Changing the exponent within the 1-$\sigma$ range leads to a small shift in the overall slope of the SN curve. Finally the parameters used in the Starburst99 program to estimate the SN response curve seen in Fig. \ref{SNresponsfunction_II} can also be changed, which in some cases results in marginally different response functions, and therefore changes in the resulting SN curves. Again, these changes are not large.

Diffusion of clusters in and out of the small region occurs in the course of time, and large initial velocity dispersions would certainly erase information about the oldest clusters forming in the solar neighbourhood. Results of a searching examination of the effect of velocity dispersions, using the model in Sect. \ref{Sec4}, are not discouraging, and in any case they may be overstating the dispersions. The simulations assume that the clusters were formed independently and an uncorrelated Gaussian distribution of velocities is used. This view may be too simple.
As shown by \citet{Piskunov2006AA} there seem to be cluster complexes consisting of 10-100 clusters that have a common dynamical origin, presumably because they were born in the same giant molecular cloud. Clusters that are members of the resulting "families" have similar ages and are located at similar distances. In the simulations in Sec. \ref{Sec4} such dynamical correlations are not included. If they were included, the likely result would be a smaller overall dispersion of the clusters, with better retention of the cluster formation history in the solar neighbourhood. The simulations shown in Sect. \ref{Sec4} may therefore be worst-case scenarios.

They also make it unlikely that the ages of nearby clusters are accidental and unrelated to the Solar System's journey through a structured Galaxy. Simulations of homogeneous star formation in the annulus including the solar circle, as displayed in Fig. \ref{SIM4FIG}c, give age distributions different from the observations in the WEBDA list. In summation the results as displayed in Fig. \ref{SIM4FIG} give confidence in the existence of a real astrophysical signal containing information about both a variable star formation history and the structure of the Galaxy near the solar circle. Reassurance also comes from the meteorite data cited in Sect. \ref{Sec3} that show almost the same average GCR rate over 500 Myr as that inferred from the cluster data.

Success with the astrophysics builds confidence in the use of the SN rates as a realistic proxy for the GCR reaching the Solar System over a long geological time-span. Results from that procedure, as in Sect. \ref{Sec6}, are sufficiently positive to suggest that, in future, geological data on GCR impacts may help to refine the astrophysics. Mutual support between astrophysics and palaeoclimatology has already occurred with the application of the Milankovitch cycles in astrochronology, as a means of achieving high resolution in geological dating over the past 65 Myr. On longer time scales, however, isotopic chronology may make geological dates more secure than stellar ages. If so, mismatches with climate might encourage a re-examination of some astrophysical data. And a foretaste of other clues for astrophysicists comes from evidence presented in Sect. \ref{Sec6} that short-lived falls in sea level recorded by seismic stratigraphy promise high-resolution dating of supernova events closest to the Earth.

\subsection{Local supernova rates and climate (Section \ref{Sec6})}
Although comparisons in this paper between astrophysical and geological data broadly endorse the link between GCR and climate found by Shaviv \citep{Shaviv2002PRL}, there are interesting differences. Approaching the subject via the open clusters and supernova rates provides a more inductive and more detailed view of astrophysical events over the past 500 Myr, for comparison with new geological information, which is itself becoming more detailed. The matches between peak ice events and maxima in SN rates (Fig. \ref{WEBDAcluster_CLIMATE}) are an example of progress in both respects. It should be mentioned that the solar luminosity was about 6 \% lower when the Sun was 500 Myr younger, so the effect of GCR peaks on cloudiness 500-350 Myr was perhaps more influential than it would be now.

The traditional use of $\delta^{18}$O data to gauge palaeoclimates is questionable over long time scales. In a recent compilation \citep{Veizer2008} the authors found that a period of 120 Myr could explain up to 70\% of the multi-million year variability of $\delta^{18}$O over the last 115 Myr. As seen here in Fig. \ref{SN_d18O_200Ma}, there are hints of cyclical variation over 200 Myr and the general trends are in broad agreement with the change in SN rate. But on the 500 Myr time scale (not shown) large spreads in $\delta^{18}$O data in each interval of time suggest that better assurance is needed about the provenance of the data, for comparing like with like (zonal versus global temperatures). Otherwise, $\delta^{18}$O is at best an approximate aid to verifying the long-term astrophysical connection to climate.

More promising is the innovation here concerning a likely link between major short-lived falls in sea level and the nearest supernovae. The proposition that intense GCR fluxes from close supernovae caused glaciations and associated eustatic regressions in sea level finds a persuasive match in the computed high temporal resolution of GCR variation based on statistics of nearby supernovae (Fig. \ref{Sealevel_SN}). Asteroidal and cometary impacts might offer a rival hypothesis, with the right sort of frequency for major marine regressions, but the very big impact in Mexico 65.5 Myr ago, which terminated the Mesozoic Era, is associated with only a minor marine regression, $<$ 25 m.

Overall, the consistency of results concerning climate provides mutual validation of the star clusters as a guide to SN events, of GCR as a climatic factor, and of geological data as a window on astrophysical events long ago.

\subsection{Local supernova rates, macro-evolution and biodiversity (Section \ref{Sec7}, first part)}
Mass extinctions that ended the Paleozoic and Mesozoic Eras, with drastic changes in the conspicuous animals, have aroused much interest in recent decades, notably concerning the sudden disappearance of the dinosaurs and many coeval marine genera at the close of the Mesozoic 65.5 Myr ago. Debate and controversy have nevertheless surrounded the facts about biodiversity - how the fossils are to be counted - with major revisions from \citet{Sepkoski1981} to \citet{Alroy2008}. The reasons for the variations in biodiversity are also matters of dispute.
Mechanisms on offer include biological processes such as spontaneous diversification with limits to species richness, and physical influences such as continental drift, extraordinary volcanic eruptions, and impacts of comets or asteroids causing mass extinctions. The role of sea level was confirmed above. Although the amplitude of the sea level changes due to tectonics has been disputed \citep{Miller2005} the shape of the long-term global cycle influencing biodiversity in marine invertebrates appears to be fairly secure, so that any revisions to the amplitude would affect only the normalizing factor applied to the data in Fig. \ref{GeneraAlroy}.
The connection with SN rates revealed in Fig. \ref{GeneraAlroy} adds to the mix of hypotheses, by providing strong evidence that Galactic cosmic rays (GCR) have influenced the course of evolution.

Since H.J. Muller in the 1920s showed that ionizing radiation causes genetic mutations \citep{Muller1927}, an obvious contribution of GCR to evolution has been well known - namely in provoking some of the mutations on which natural selection works. The importance of GCR in this respect remains uncertain because other causes of mutagenesis include solar protons, radioactivity, environmental chemicals, thermal shock and transcription errors. Natural repair mechanisms that organisms possess may be better adapted to continuous hazards like GCR than to rare events like thermal shock.

On the other hand, GCR seem to exert a strong though indirect evolutionary influence by varying the climate. A persistently warm climate tends to reduce global biodiversity, because there is little motivation to evolve, dominant species keep others in check, and there is less variety in habitats and living conditions between the tropics and the polar regions. In a cold and variable climate, on the other hand, the dominant species are stressed and this gives opportunities to other species, in accordance with the intermediate disturbance hypothesis that traces back to \citet{Grime1973}. Cold conditions also provide a greater variety of habitats and living conditions.
There is therefore no reason to disbelieve the SN-biodiversity link. What is perhaps surprising is how much of the variation in the counts of marine invertebrate genera is accounted for by astrophysics (SN) plus tectonics (sea level). The message of Fig. \ref{GeneraAlroy} may be that here is evidence concerning evolution analogous to plate tectonics in geology, imposing an overarching simplicity while leaving mismatches to explain and an infinity of detail to explore.

Conspicuous among the topics worth re-examining is the largest mass extinction in the past 500 Myr, the Permo-Triassic event of 251 Myr ago, when the level of loss of marine species reached 80-96\% and two-thirds of tetrapod families went extinct on land along with eight orders of insects \citep{Sahney2008}. The main hypotheses competing to explain it invoke volcanism in Siberia or impacts by asteroids or comets, and part of the marine loss can be attributed to an exceptionally low long-term sea level (Fig. \ref{GeneraSealevel}). But the Permo-Triassic event also coincided with the largest decrease in SN rates from one 8 Myr bin to the next in the whole 500 Myr period, as seen at -252 Myr in Fig. \ref{WEBDAcluster_CLIMATE}. That was when Solar System had left the very active Norma spiral arm. Fatal consequences would ensue for marine life if a rapid warming led to nutrient exhaustion (see below) occurring too quickly for species to adapt.
Before the culminating Permo-Triassic event, the earlier Guadalupian-Lopingian mass extinction occurred about 265 Myr ago. By contrast, this was linked with global cooling that interrupted a mid-Permian warming, and with a large increase in $\delta^{13}$C in the interval 265-260 Myr. \citet{Isozaki2007} point out that prominent victims of this extinction were warm-water fauna including gigantic bivalves and rugose corals. The full sequence of warming, cooling and precipitate re-warming in the latter part of the Permian period conforms very well with the variations in SN rates.

\subsection{Local supernova rates: $\delta^{13}$C and bio-productivity (Section \ref{Sec7}, second part)}
The working hypothesis mentioned above, that in SN-induced glacial conditions primary productivity increases and leads to drawdown of CO$_{2}$ and $^{12}$C sequestration, is compatible with the empirical result in Fig. \ref{SN_delta_C13}. A reality check comes from remarkable peaks in reef carbonate production 430, 380, 290, 150 and 20 Myr ago (\citet{Kiessling2005}, Supp. Fig. 3). All of these peaks coincide with high SN rates, except for the maximum in the SN rate 150 Myr ago, which was relatively weak. A link between SN rates and productivity is therefore clear, but in view of the complexity of the carbon cycle and the many factors governing organic carbon burial or entrapment, and hence $\delta^{13}$C, it might be unwise to conclude that the mechanism is as simple as stated.

On the other hand, investigators have often noted the coincidence of $\delta^{13}$C excursions with climatic episodes. For example, high values of $\delta^{13}$C in the Silurian Period are linked \citep{Stricanne2006} to episodes of aridity in the tropics, which are often associated with polar glaciation, although that is questioned in this particular case. Over a longer timespan, Cambrian to Permian, large increases in $\delta^{13}$C were rare during hothouse climates but common during cool periods \citep{Saltzman2005}.
The productivity of the oceans is limited by the supply of nitrogen in the near-surface waters, which is turn is limited by the availability of iron needed by oceanic nitrogen-fixing cyanobacteria. During recent glacial periods, iron delivered by windblown dust was at least an order of magnitude higher than during interglacial periods \citep{Falkowski1998} and at the Last Glacial Maximum $\approx$20 kyr ago organic carbon concentrations in the sediment of the eastern equatorial Pacific were twice those of the past 11 kyr (Holocene) \citep{Pichevin2009Natur}.

Occurrences of anoxia (regional exhaustion of oxygen) have been offered as an explanation for the variability in $\delta^{13}$C in cold periods, with episodic organic carbon burial sustained by positive feedbacks between productivity and anoxia, whilst more stable $\delta^{13}$C regimes in warm periods, which can last $>$10 Myr, may tell of a nitrogen-limited ocean in which anoxia leads to increased denitrification \citep{Saltzman2005}. In addition, exhaustion of nutrients can curb productivity in cold periods, causing variations in $\delta^{13}$C. A related matter for consideration is the likely contribution to $\delta^{13}$C variations of the short-lived glacial episodes apparently linked to the nearest supernovae (Sect. \ref{Sec6}).
For nearly all of the geological record of the past 500 million years, long-term variations of $\delta^{13}$C from organic and carbonate sources have gone up and down more or less in step. There is, however, a marked divergence in the most recent part of the record that may be connected with the emergence of C4 terrestrial plants, which have an extra carbon atom in their first photosynthetic molecules and markedly higher $\delta^{13}$C compared with the commoner C3 plants \citep{Osborne2006}. In Fig. \ref{SN_delta_C13} there is a good match between organic $\delta^{13}$C data to rising SN rates in the past 20 Myr, but carbonate $\delta^{13}$C (not shown) diminished.

These are only hints about how the empirical evidence about $\delta^{13}$C may be explained, and future theoretical inputs will no doubt go far beyond the working hypothesis of drawdown and burial. The challenge is similar to the case of biodiversity, for which many processes of evolution and extinction exist and yet a dominant role for astrophysics is discernible. In the same spirit, specialists are invited to make of what they can of the $\delta^{13}$C link to the SN rate seen in Fig. \ref{SN_delta_C13}.
Finally, biological activity from phytoplankton is a source of sulphur compounds in the atmosphere and therefore also of cloud condensation nuclei \citep{Charlson1987Natur}. So a significant increase in biological productivity associated with a cooling climate would give positive feedback on cloud formation and thereby encourage further cooling. Conversely, nutrient limitation due to either excessive productivity or weak circulation could exert a warming effect by reducing cloud formation.

Other processes might in principle counter a cooling of the climate by negative feedback. For example, if a cooling reduces the loss of CO2 to geochemical weathering, that could lead to a buildup of CO2 if other sinks and sources of CO2 remain constant, and so dampen or reverse the cooling \citep{Donnadieu2009}. The net impact of such processes will eventually have to be resolved by estimating the forcing of the individual processes and incorporating them in a multidisciplinary numerical model beyond the scope of this paper.

\subsection{Caveats}
The present results are no better than the data on which they are built, and the uncertainty of the data gets larger as one reaches further into the past. Estimated variations in SN rates are better determined for 250 - 0 Myr ago than for 500 - 250 Myr ago, because there are fewer clusters of old ages and because phase mixing tends to erase part of the memory of the birthplaces of the open clusters. On the other hand, consistencies in the comparisons with geological data lend support to the estimated SN rates, even in the earlier period. Without direct terrestrial records of the GCR variation on long time scales, the highly fluctuating flux due to nearby SNs remains a matter of numerical modelling, but again the geological record of sudden drops in sea level appears to support the analysis. It remains to be seen whether future improvements in astrophysical and geological data will show them continuing to mesh in the ways suggested here.

\section{CONCLUSION}
\label{conc}
The impact of stellar and interstellar processes on the Earth has been addressed many times but, with the exception of \citet{Shaviv2002PRL}, the speculations about episodic effects of e.g. a closer-than-usual SN \citep{Fields1999NewA} or variations in the interstellar medium \citep{Frisch2000AmSci} have demonstrated no persistent influence on life and climate. The present paper has aimed at revealing the integrated effect of varying GCR flux on the Earth during the past 500 Myr by reconstructing the SN rate during this period, and comparing it with geophysical and palaeobiological data.

Using the WEBDA database on open cluster formation, the SN rate in the solar neighbourhood ( $<$ 0.85 Kpc) was constructed, and the variation was shown to be a superposition of two distinct features originating either inside or outside the solar circle. Outside the solar circle are four clear maxima in the SN reconstruction at 372.6, 269.1, 140.6, and 17.3 Myr before present, which are interpreted as the Solar System's encounters with four spiral arms, leading to a relative angular frequency of the Solar System $\Omega_0-\Omega_P$ = 13.0 $\pm $ 0.9 $\textrm{km s}^{-1} \textrm{kpc}^{-1}$ (or $\Omega_P/\Omega_0 = 0.57 \pm 0.5$). Inside the solar circle the pattern is not simple. In checking that the velocity dispersions of clusters do not erase the star formation history, a model simulating cluster motions in and out of the solar neighbourhood gave satisfactory results, even without taking account of the beneficial effect of the birth of many clusters in families with relatively low dispersions.

In addition it was shown that multiple SN sources generate a highly fluctuating GCR signal and that the nearest SNs (closer than $\approx$ 300 pc) produce clear spikes in the GCR flux. Those provided the first link to the terrestrial climate noted here, with previously unexplained short-lived falls in sea level accounted for by sudden but brief glaciations caused by the nearest SNs, as in Fig. \ref{Sealevel_SN}.
The match of longer-term climate to SN rates, seen in Figs. \ref{WEBDAcluster_CLIMATE} and \ref{SN_d18O_200Ma} leaves room only for relatively small or brief climatic influences of all the tectonic, volcanic, and other processes discussed in this connection. If the dominant role of the Galaxy in the terrestrial environment is further validated by ongoing studies, cosmic and terrestrial, it promises to simplify Phanerozoic climatology.

As for the palaeobiology, remarkable connections to the long-term histories of life and the carbon cycle have shown up unbidden (Figs. \ref{GeneraAlroy}, \ref{GEOCO2}, \ref{SN_delta_C13}). Biodiversity, CO$_2$ and $\delta^{13}$C all appear so highly sensitive to supernovae in our Galactic neighbourhood that the biosphere seems to contain a reflection of the sky.

\section*{Acknowledgments}
This research has made use of the WEBDA database (http://www.univie.ac.at/webda/), operated at the Institute for Astronomy of the University of Vienna. The author also thanks the many geoscientists who have provided data and helpful comments, and Nigel Calder for general discussions regarding this work.

\bibliographystyle{mn2e} %
\bibliography{HSVref}    
\end{document}